\documentclass[prc,aps,groupedaddress,preprintnumbers,onecolumn]{revtex4}

\usepackage{graphicx} 
\usepackage{amssymb}
\usepackage{epstopdf}
\usepackage{hyperref}
\usepackage{epsfig}
\usepackage{multirow}
\usepackage{verbatim}
\usepackage{url}

\newcommand{\nue}{\ensuremath{\nu_{e}}}
\newcommand{\nuebar}{\ensuremath{\overline{\nu}_{e}}}
\newcommand{\antinue}{\ensuremath{\overline{\nu}_{e}}}
\begin{document}

\title{Experimental Parameters for a Reactor Antineutrino Experiment at Very Short Baselines}

\author{K.M.~Heeger, }
\author{M.N.~Tobin}
\affiliation
        {\it Department of Physics,
University of Wisconsin, Madison, WI 53706, USA}
\author{B.R.~Littlejohn}
\email{littlebh@uc.edu}
\affiliation
        {\it Physics Department,
University of Cincinnati, Cincinnati, OH 45221, USA}
\author{H.P.~Mumm}
\affiliation
        {\it National Institute of Standards and Technology, Gaithersburg, MD 20899, USA}
\date{\today}


\begin{abstract}
{Reactor antineutrinos are used to study neutrino oscillation, search for signatures of non-standard neutrino interactions, and to monitor reactor operation for safeguard applications. The flux and energy spectrum of reactor antineutrinos can be predicted from the decays of the nuclear fission products. A comparison of recent reactor calculations with past  measurements at baselines of 10-100\,m  suggests a 5.7\% deficit. Precision measurements  of reactor antineutrinos at very short baselines $\mathcal{O}$(1-10\,m) can be used to probe this anomaly and search for possible oscillations into sterile neutrino species. This paper studies the experimental requirements for a new reactor antineutrino measurement at very short baselines and calculates the sensitivity of various scenarios. We conclude that an experiment at a typical research reactor provides 5$\sigma$ discovery potential for the favored oscillation parameter space with 3 years of data collection.}
\end{abstract}

\maketitle

\section{Introduction}
\label{sec:introduction}

Recent calculations of the predicted $\nuebar$ flux from reactors compared to past  measurements at baselines between 10-100\,m have revealed an apparent deficit of about 5.7\%~\cite{Anomaly2}. This discrepancy, the {\em ``reactor anomaly''}, can be interpreted as a sign of new physics  or point to an issue with the reactor flux calculations ~\cite{Hayes}. Independent calculations have verified the reactor $\antinue$ flux predictions~\cite{HuberAnomaly} but subsequent calculations based on a full \textit{ab initio} prediction using newly available nuclear data~\cite{AnomalyFix} reduce the size of the discrepancy. It has been suggested that the ``{\em reactor anomaly}''  may be the signature of additional sterile neutrino states with mass splittings of the order of $\sim 1{\rm eV}^2$ and oscillation lengths of $\mathcal{O}$(3\,m)~\cite{Anomaly1}.  Other anomalies in neutrino physics including the observation of apparent $\nu_e$ and $\nuebar$ appearance at similar mass-squared splittings in accelerator experiments~\cite{LSND, MB}, deficits in observed events from high-intensity $\nu_e$ sources used to calibrate solar neutrino detectors~\cite{Gallium}, and the preference for more than three relativistic species in astrophysical surveys ~\cite{Astro} add to the puzzle.

 Additional data will soon be provided by the currently-operating km-scale reactor experiments, Daya Bay, Double Chooz, and RENO~\cite{DB,DC,RENO}.   While highly precise, these experiments cannot resolve the short oscillation lengths associated with eV$^2$ mass splittings; at these distances and with a finite detector energy resolution, the oscillation effect from potential sterile states averages to yield an effective rate deficit.  Moreover, these measurements will eventually be limited by the understanding of the interference of multiple reactor cores, the presence of oscillation effects from other mass-squared splittings, and by the inability to take background data free of reactor $\nuebar$. A new experiment at very short ($<$10~m) baselines with a single core in a controlled research environment where backgrounds can be measured independently is needed to fully disentangle reactor flux and spectrum prediction uncertainties from sterile neutrino oscillation effects, or other signs of new physics. 

The paper is organized as follows: Section~\ref{sec:parameters} summarizes both the experimental detector and reactor parameters under consideration. Section~\ref{sec:sig} introduces the characteristic signature of neutrino oscillations and defines the $\chi^2$ analysis used to calculate experimental sensitivity. A nominal, generic reactor-detector arrangement referred to as the ``default experiment'' is defined for systematic studies and comparison.
Sections~\ref{sec:reac},~\ref{sec:facil},~\ref{sec:Bkgs}, and~\ref{sec:detect} present a discussion of reactor, facility, background, and detector parameters and examine the impact on the overall sensitivity of the experiment.
Section~\ref{sec:summary} summarizes the sensitivity and discovery potential of a new reactor experiment and discusses its optimization.

\section{Experimental Parameters}
\label{sec:parameters}

A number of experiments have been proposed to address the reactor anomaly by making a precise measurement of the reactor flux and spectrum at very short baselines~\cite{HeegerTalk, LasserreTalk, BowdenTalk, Serebrov}.  The proposed experimental configurations are site-specific and reflect the infrastructure and logistical constraints at each site. This work studies generic experimental parameters  that determine the ultimate sensitivity of a reactor experiment at very short baselines, and helps guide the optimization of a new experiment aimed at a definitive sterile neutrino oscillation search. 


\subsection{Reactor Parameters}

Antineutrinos from the reactor core are used as the flavor-pure $\nuebar$ source. The relevant reactor reactor core parameters are the following:

I. {\em Reactor power:} Each nuclear fission initiates the release of a known amount of energy along with several antineutrinos.  The reactor $\nuebar$ flux is proportional to the thermal power output of the reactor modulo corrections for the isotopic fuel composition.  High operating power combined with high reactor up-time maximize the event statistics. Most commercial reactors, but not all research reactors, operate close to their licensed power.

II. {\em Fuel type:} Most nuclear reactors utilize uranium-based fuel containing a mixture of $^{238}$U and $^{235}$U, with the latter providing the majority of total fissions and thermal power.  The two most common classes of reactor fuels, commercial and highly-enriched, differ primarily in the total percentage enrichment of $^{235}$U.  Highly-enriched uranium (HEU) fuel, commonly utilized in research reactors, contains upwards of 90\% $^{235}$U, with fission fractions dominated by $^{235}$U at all points in the fuel cycle.  Commercial fuel generally contains less than 6\% $^{235}$U.  While $^{235}$U fission comprises the majority of all fissions in these cores, a significant fraction (30-50\%, depending on the fuel burnup) is contributed by $^{238}$U and by $^{239}$Pu and $^{241}$Pu accumulated during the fuel cycle~\cite{Kopeikin}.  These isotopes produce different proportions of fission products, resulting ultimately in a difference in the total number, energy spectrum, and spectral uncertainty of produced $\overline{\nu}_e$.

III. {\em Duty cycle:} The duty cycle is given by the ratio of the power cycle to the subsequent shutdown periods for refueling or maintenance. The duty cycle influences the statistical power of an experiment in a straightforward way.  Therefore, throughout this paper, we present results only in terms of detector livetime.  Importantly, reactor down-time can be used for determining the shape and position distribution of backgrounds~\cite{DCOff}.

IV. {\em Core dimensions:} Antineutrinos are produced  throughout the active reactor core and emitted isotropically.  The core's finite dimensions cause a spread in the neutrino path lengths between the reactor core and detector that washes out the observable oscillation in the detector.

As examples, we use parameters from three research reactors in the US, the High Flux Isotope Reactor (HFIR) at Oak Ridge National Laboratory (ORNL)~\cite{HFIR}, the National Bureau of Standards Reactor (NBSR) at National Institute of Standards and Technology (NIST)~\cite{NBSR}, and the Advanced Test Reactor (ATR) at Idaho National Laboratory (INL)~\cite{ATR}. The Institut-Laue Langevin (ILL) reactor in France and the commercial San Onofre Nuclear Generating Station (SONGS) in California are included for reference~\cite{ILL, SONGS}.




 \subsection{Facility Parameters Parameters}
 
The reactor facility provides constraints on the location of the detector, its size, and distance to the core. The following parameters determine the experiment's design and sensitivity: 

I. {\em Experimental space and detector volume:}  Detector location, geometry, and dimensions are limited by the available space near the reactor core. Typically, biological shielding limits the distance of closest approach to the reactor while the cross-sectional area of the detector and the maximum radial distance are constrained by the layout of the reactor facility, floor-loading limits, and the availability of floor space not occupied by other experiments, reactor equipment, or detector shielding.  To maximize event statistics, a large cross-sectional area at close radial distances is desirable.  

II. {\em Detector distance to core:} The closest distance to the reactor determines the overall $\antinue$ flux seen by the detector as well as the oscillation wavelength probed. For the following discussion we define this distance of closest approach, $r_{min}$, to be from the center of the reactor core to the closest point in the active detector target.  

Throughout this discussion we explore a range of detector sizes and closest distances reflecting the expected space available at US facilities.

 \subsection{Detector Parameters}
 
Intrinsic detector characteristics determine the observed event distributions:

I. {\em Fiducial volume and target mass:} The observed number of $\nuebar$ interactions in the detector scales with the number of target protons in the detector and sets the statistical precision of the experiment.  

II. {\em Detection efficiency:} The detection efficiency directly scales the detected number of events. For fixed signal-to-background, S/B, changes in the detection efficiency have the same effect as increasing the detector volume or target mass. 

III. {\em Position and energy resolution:} An unambiguous demonstration of neutrino oscillation requires the observation of oscillations in energy and distance, the characteristic $L/E$ dependence.  Good position and energy resolution are necessary to maximize the experimental sensitivity.

 \subsection{Backgrounds}
 
A reactor experiment at very short baseline requires the operation of a detector on the surface under minimal overburden. Muon and cosmic ray-induced backgrounds as well as backgrounds from the reactor are important considerations.  Both the magnitude and shape of backgrounds are expected to be site-specific.  Reactor-correlated backgrounds are affected by the reactor-detector distance, the reactor power, and the amount of shielding material between the reactor and detector.  Cosmic background rates are dependent on the total overburden provided by the facility and the mass distribution near the detector location.  Backgrounds can be mitigated with active or passive shielding, as well as with background rejection techniques, such as pulse-shape discrimination or detector segmentation.  Small but irreducible backgrounds similar in energy spectrum to the signal may also be provided by spent nuclear fuel nearby the reactor~\cite{SNF}.  Detailed background studies are not presented for any particular reactor sites in this paper; instead, reasonable background models are constructed based on experience in previous experiments~\cite{ILL, Bowden, Lasserre, Bryce}.  Specifically, we consider the following parameters:

I. {\em Signal-to-background ratio (S/B):} The overall S/B ratio is a measure of the total magnitude of time-coincident backgrounds and accidental background in the delayed coincidence window of the inverse beta-decay reaction at all energies and positions. 

II. {\em Background spectral shape:}  The backgrounds' energy dependence will determine the measured spectral shape of the observed events. Energy-dependent background subtraction and/or fitting will be important for the analysis of the energy-dependent oscillation signature. 

III. {\em Background position distribution:}  In the vicinity of a reactor significant spatial variations of backgrounds are expected. In particular, fast neutrons can scatter from surrounding materials, building structures, and even other experiments. Measurement of the observed event rate as a function of position through detector movement or segmentation will be critical for understanding local background variations on the meter-scale.

\section{Experimental Signature}
\label{sec:sig}

\subsection{Oscillations of Reactor Antineutrinos}
Antineutrinos from reactors are produced as flavor-pure $\antinue$ in the decays of the  neutron-rich fission products in the reactor fuel. 
More than 99.9\% of all $\nue$ emitted from commercial reactors are produced within the decay chains of  four isotopes, $^{235}$U, $^{238}$U, $^{239}$Pu, and $^{241}$Pu. The thermal heat released in the nuclear decays is proportional to the number of emitted $\antinue$ and is thus a measure of the flux of expected antineutrinos.  The spectrum of reactor antineutrinos detected via inverse beta decay has a mean energy of about 4\, MeV and extends up to roughly 10\, MeV. 

Neutrinos and antineutrinos are produced as a linear combination of mass eigenstates and their flavor is associated with the accompanying lepton. Due the difference in the mass eigenstates the flavor of observed neutrinos oscillates as a function of baseline and energy. For the three active neutrino states the neutrino mixing parameters are well measured in atmospheric, solar, reactor, and accelerator based experiments. Reactor $\nuebar$ disappearance over baselines of 1-2\, km and 180\, km has been observed.  The oscillation probability can be parameterized in terms of the mass splitting $\Delta m^2_{ij}$ and the mixing angle $\theta_{ij}$ between the ith and jth mass eigenstate. Additional sterile neutrino mass states with $\Delta m^2 \sim 1$\, eV$^2$ beyond the 3 active neutrinos would yield  an oscillation effect over meter-long baselines with survival probability described by 

\begin{eqnarray}
\label{eq:Detected2}
P_{sur}(E,\vec{L}) \simeq 1-\sin^22\theta_{ee}\sin^2\left(\frac{1.27 \Delta m^2_{41}| \vec{L}-\vec{r} |}{E}\right),
\end{eqnarray}
 with oscillation parameters $\Delta m^2_{41}$ and $\sin^22\theta_{ee}$. 

\begin{figure}[htb!pb]
\centering
\includegraphics[trim=0.1cm 0.1cm 0.1cm 1.4cm, clip=true, width=0.95\textwidth]{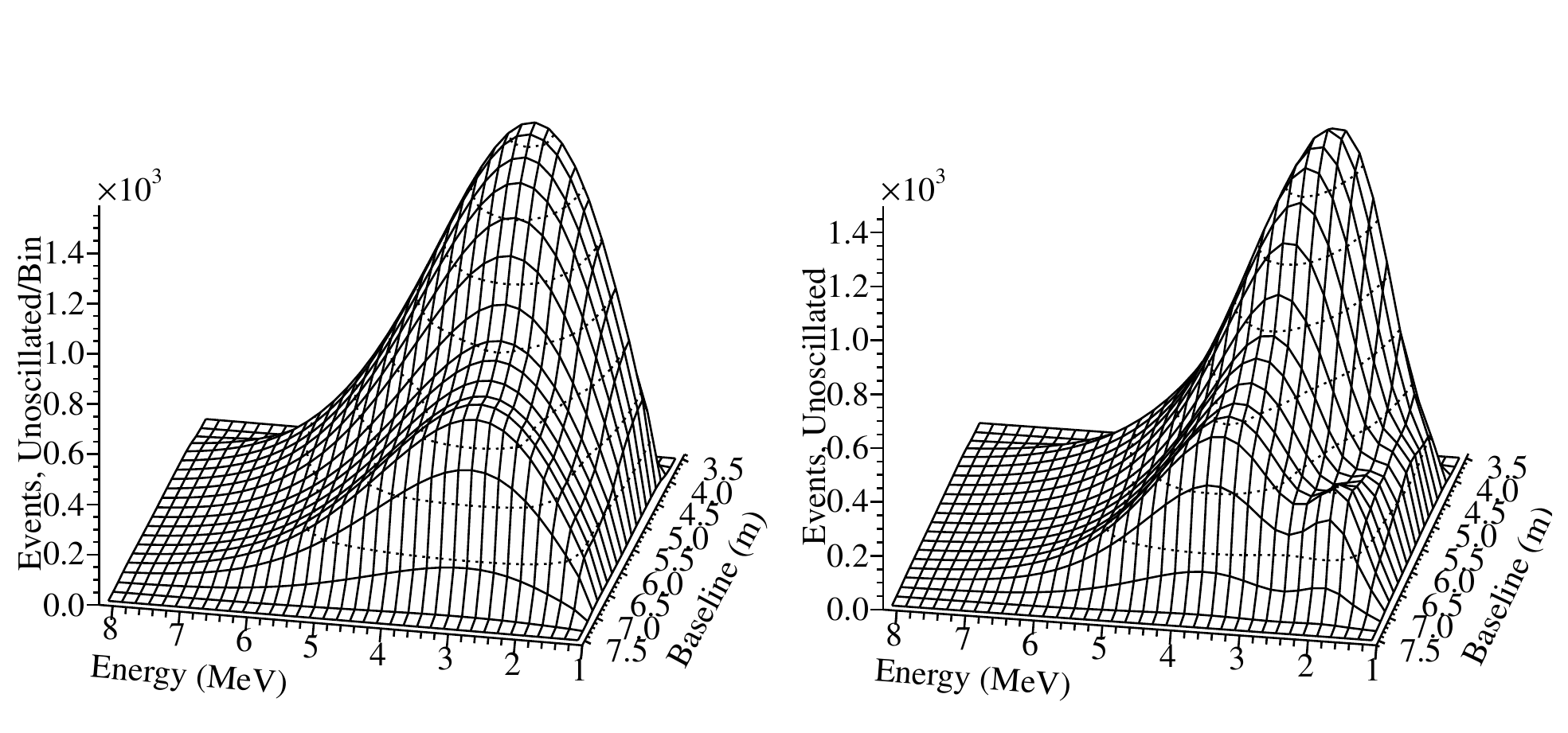}
\caption[]{Unoscillated (left) versus oscillated (right) detected $\nuebar$ spectra at ($\Delta m^2=1.8$\,eV$^2$, $\sin^22\theta_{ee}=0.5$) for a reactor $\nuebar$ detector with realistic parameters as described in Table~\ref{tab:DefaultChar}.  Exaggerated oscillation values are chosen for illustrative purposes.  The change in spectral shape from baseline to baseline is a key signature of neutrino oscillations.}
\label{fig:Oscillations}
\end{figure}

\begin{figure}[htpb]
\centering
\includegraphics[width=0.6\textwidth]{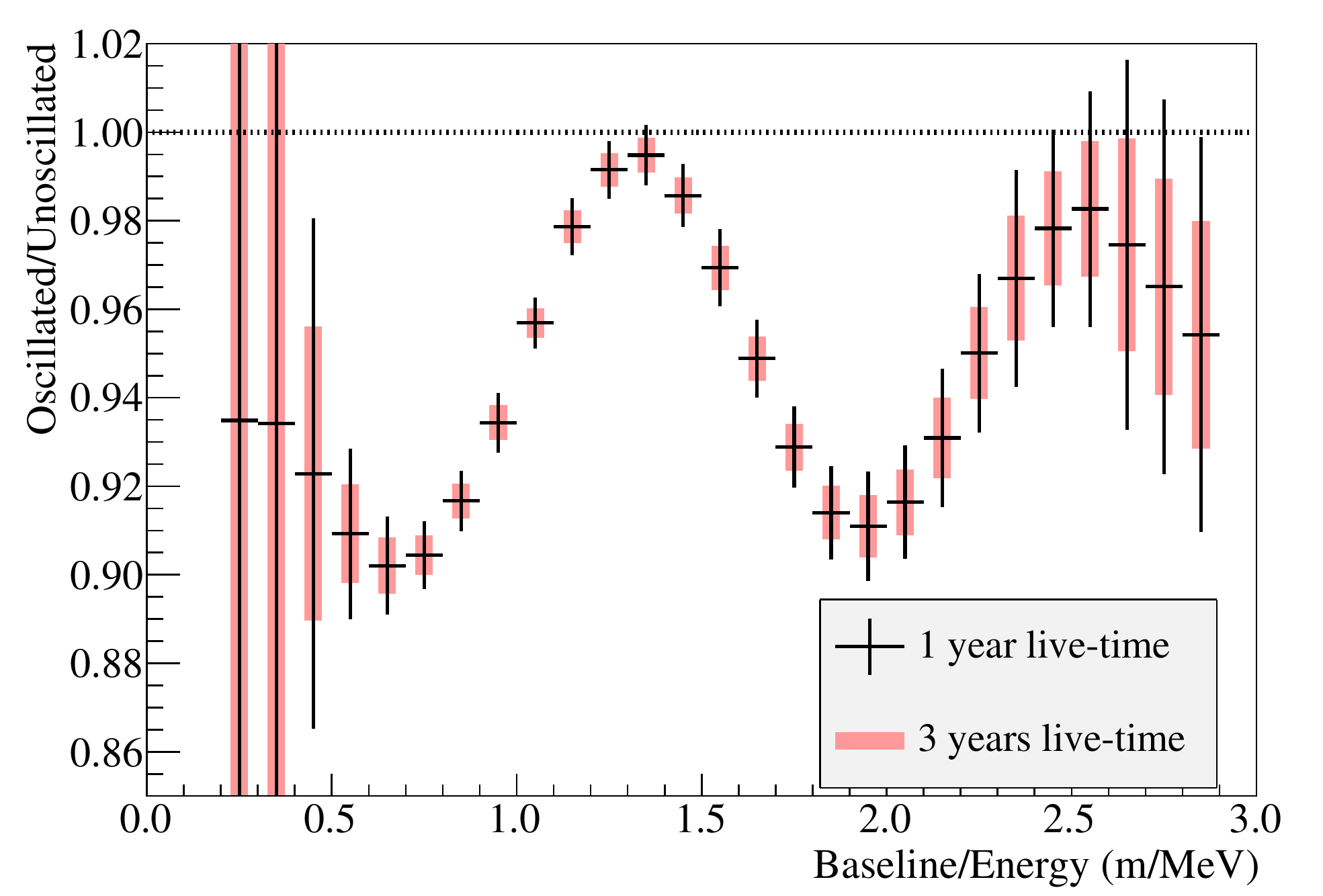}
\caption[]{Ratio of the oscillated to unoscillated spectrum as a function of L/E  for ($\Delta m^2$ = 1.8 eV$^2$, $\sin^22\theta_{ee}$=0.1).  
 and the nominal experimental arrangement described in \ref{sec:nominalexperiment}. Error bars are purely statistical.}
\label{fig:LEOsc}
\end{figure}

Figure~\ref{fig:Oscillations} illustrates the oscillation effect in baseline and energy for  ($\Delta m^2$=1.8~eV$^2$,~$\sin^22\theta_{ee}$=0.5). The characteristic L/E oscillation is pictured in Figure~\ref{fig:LEOsc} for the sterile neutrino oscillation parameters (1.8~eV$^2$,0.1) preferred in global fits~\cite{Anomaly1} and the default experimental arrangement described below.

\subsection{A Nominal Reactor Antineutrino Experiment at Very Short Baselines}
\label{sec:nominalexperiment}

Reactor neutrino experiments typically utilize the inverse beta decay reaction $\antinue+p \rightarrow e^+ + n$ (IBD) with a threshold of roughly 1.8 MeV to measure the flux and energy spectrum of reactor $\nuebar$. Liquid scintillator detectors provide a proton-rich target with high detection efficiency and good energy resolution. Rejection of backgrounds is achieved by means of the delayed coincidence signature and correlating cosmic rays with muons-induced events. Energy resolutions of around 8\% and total detection efficiencies of 75-85\% have been obtained in recent large-scale underground experiments \cite{DB, DC}. For smaller detectors developed for early short-baseline $\nuebar$ experiments and more recent reactor monitoring purposes, energy resolutions of 10-20\% and efficiencies of 10-50\%  have been reported \cite{Bowden, Bugey, Rovno, Rovno2}.   For the studies presented in this paper a nominal experimental arrangement of $\mathcal{O}$(m$^3$)-sized detector at distances of $\mathcal{O}(10\,m)$ from an average research reactor is considered. The specific parameters assumed for the nominal experiment are listed in Table~\ref{tab:DefaultChar}.

\begin{table}[!htb]
  \noindent\makebox[\textwidth]{%
    \begin{tabular}{l | l | l | l | l} \hline
      \multicolumn{2}{c|}{Parameter} & Value & Comment & Reference \\ \hline
      \multirow{4}{*}{Reactor} & Power & 20 MW & NIST-like & \cite{NISTschedule}\\
      & Shape & cylindrical & NIST-like & \cite{NISTschedule} \\
      & Size & 0.5~m radius, half-height & NIST-like & \cite{NISTschedule} \\
      & Fuel & HEU & Research reactor fuel type & \cite{NISTschedule, HFIRschedule, ATRschedule}\\ \hline
      \multirow{5}{*}{Detector} & Dimensions & 1$\times$1$\times$3~m & 3 meters of available baseline & - \\
      & Efficiency & 30\% & In range of SBL exps. (10-50\%) & \cite{Bugey, Rovno, Bowden} \\  
      & Proton density & 6.39$\times$10$^{28}\frac{p}{m^3}$ & From Daya Bay GdLS & \cite{TDR} \\
      & Position resolution & 15~cm & Daya Bay-like & \cite{Bryce}  \\
      & Energy resolution & 10\%/$\sqrt{E}$ & Daya Bay-like & \cite{DBNIM} \\ \hline
      \multirow{4}{*}{Other} & Run Time & 1 year live-time & -  & - \\ 
      & Closest distance & 4~m & NIST-like & - \\
      & S:B ratio & 1:1 & In range of SBL exps. (1-25) & \cite{Bugey, Bowden, ILL}\\
      & Background shape & 1/E$^2$ & Similar to SBL experiments & \cite{ILL, Bowden, Bryce} \\
      \hline
    \end{tabular}}
  \caption{Nominal experimental parameters used for the sensitivity calculations presented in this paper.}
  \label{tab:DefaultChar}
\end{table}

\subsection{Experimental Sensitivity and Discovery Potential}
\label{subsec:chisquare}

The sensitivity of a reactor experiment to neutrino oscillations is evaluated by comparing the detected energy spectrum to the expected one in the absence of neutrino oscillations. A radially  extended detector with position resolution allows a comparison as a function of baseline. A $\chi^2$ test is used to test the hypothesis of no-oscillation and for parameter estimation in $\Delta m^2_{41}$ and $\sin^22\theta_{ee}$. Observed $\nuebar$ events are binned in energy $E$ with index $i$ and in distance between the core and the point of detection, $|\vec{L}|$, along index $j$. The expected unoscillated  number of events per bin, $T_{ij}$, is given by 

\begin{eqnarray}
\label{eq:Detected}
T(E,\vec{L}) = \frac{N_p \epsilon}{4 \pi} \int  \frac{\sigma(E) S(E)}{| \vec{L}-\vec{r} |^2} d\vec{r},
\end{eqnarray}

with $N_p$ as the number of target protons, the detection efficiency $\epsilon$, energy $E$, point vector $\vec{r}$ between the core center and $\overline{\nu}_e$ production point in core, vector $\vec{L}$ between the core center and $\overline{\nu}_e$ detection point in the detector, $\nuebar$ energy spectrum $S(E)$, and inverse beta cross-section $\sigma(E)$. $M_{ij}$ is the expected number $T_{ij}$ with backgrounds added in the presence sterile neutrino oscillations as described by Equation~\ref{eq:Detected2}.  Both $M_{ij}$ and $T_{ij}$ are subject to gaussian position and energy resolution smearing according to the values given in Table~\ref{tab:DefaultChar}.  The $\chi^2$ function is defined as:

\begin{eqnarray}
\label{eq:chi2}
\displaystyle
\chi^2 = \sum_{i,j}\frac{\left[M_{ij} - (\alpha + \alpha_e^i + \alpha_r^j)T_{ij} - (1+\alpha_b)B_{ij} \right]^2 }{T_{ij}+(\sigma_{b2b}B_{ij})^2}  
+\displaystyle \frac{\alpha^2}{\sigma^2} + \displaystyle\sum_{j}\left(\frac{\alpha_r^j}{\sigma_r}\right)^2 + \displaystyle\sum_{i}\left(\frac{\alpha_e^i}{\sigma_e^i}\right)^2 + \frac{\alpha_b^2}{\sigma_b^2}.
 \end{eqnarray}
 
The $\chi^2$ sums over 17 prompt energy and 19 position bins in the range of [0.8,7.6]\, MeV and [3.2, 7.4]\, m, with bin withs of 0.4~MeV and 0.2~m respectively. The bin widths are comparable to their respective modeled detector resolutions (10\%/$\sqrt{E}$, 0.15~m).  The sum is minimized with respect to $\theta_{ee}$, $\Delta m^2_{41}$ and to the nuisance parameters \{$\alpha$, $\alpha_r^j$, $\alpha_e^i$, $\alpha_b$\}, as described in~\cite{PDG}.  The parameter $\alpha$ allows the signal normalization to vary within the bounds of its associated uncertainty $\sigma$ to account for uncertainties in the absolute reactor $\nuebar$ normalization and absolute detection efficiency.  The 100\% error in $\sigma$ ensures a floating overall normalization, meaning sensitivity to oscillation is only given by spectral distortions.  

The parameters $\alpha_e^i$ account for the the uncertainty in the reactor $\overline{\nu}_e$ spectrum from reactor flux predictions and from previous experimental measurements, as well as detector systematics uncorrelated between energy bins. These parameters allow position bins at one energy to fluctuate together, independently of position bins at any other energy.  These correlated fluctuations are limited by $\sigma^i_e$, which vary as a function of energy as described in~\cite{HuberAnomaly}.  Modelling of these uncorrelated errors in energy is of particular importance, as they limit the power of a pure energy-based oscillation analysis.  This will be discussed further in Section~\ref{subsec:Versus}.

The position spectrum parameters $\alpha_r^j$ allow correlated fluctuations with position, rather than energy, in order to incorporate the effects of relative efficiency differences and uncertainties between position bins.  In contrast to the energy spectrum uncertainties, relative efficiency uncertainties between position bins, given a value $\sigma_r$=0.5\% for all bins, should be smaller, as they are easier to characterize via detector simulation and calibration. 

Backgrounds $B_{ij}$ are estimated with a flat position dependence and an energy spectrum that falls with energy as $1/E^2$, generally mirroring the background shape reported by some previous short-baseline experiments~\cite{ILL, Bowden, Lasserre, Bryce}.  The background normalization is allowed to fluctuate similarly to the signal normalization by incorporating the background nuisance parameter $\alpha_b$, with an associated systematic uncertainty of $\sigma_b$=10\%.

Uncorrelated uncertainties in the background energy and position spectrum should be very well-characterized for a successful experiment.  The sources of uncertainties in position spectrum and in energy spectrum for backgrounds are not likely to be mostly decoupled, as is the case for the signal position and energy spectrum uncertainties.  Without the specifics of these correlations, the most conservative way to incorporate such an uncertainty into the $\chi^2$ analysis is to provide an additional independent nuisance parameter for every bin with an associated uncertainty $\sigma_{b2b}$ that reflects the precision of any background spectral measurements.  Conservative consideration of these uncertainties can be more simply achieved by adding the effect of $\sigma_{b2b}$ to the denominator of the $\chi^2$.  For this study, $\sigma_{b2b}$ is given a default value of 0.5\%.

\begin{figure}[htb!]
\centering
\includegraphics[width=0.6\textwidth]{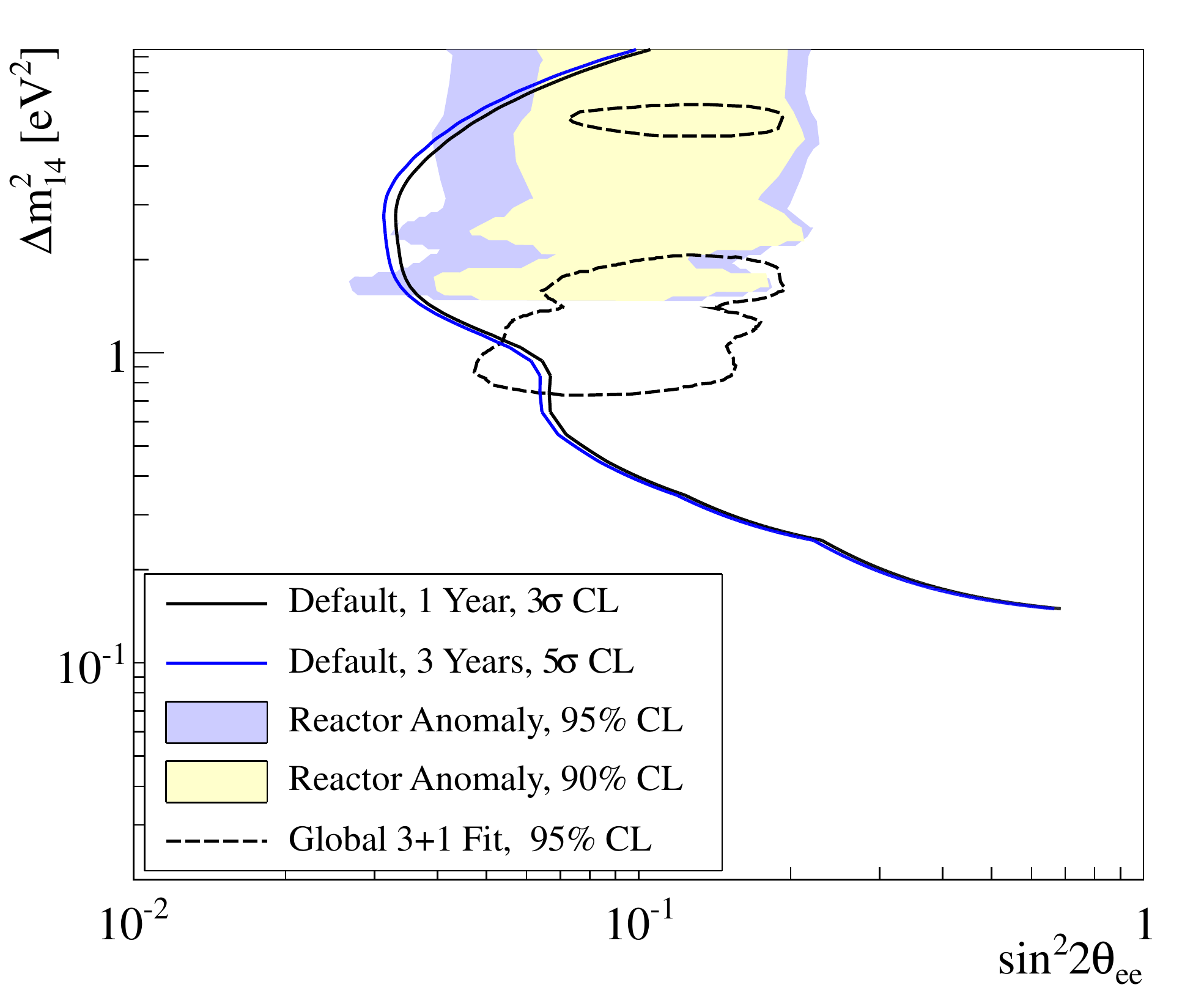}
\caption[]{Discovery potential to neutrino oscillations at 3- and 5$\sigma$ for the nominal experimental arrangement described by Table~\ref{tab:DefaultChar}.  The 3$\sigma$ contour corresponds to one year of detector live-time while the 5$\sigma$ contour is shown for three years of detector live-time. Also pictured are the best-fit parameter spaces for the reactor antineutrino anomaly~\cite{Anomaly1} and for the 3+1 global fit to all relevant accelerator, source, and reactor data given in~\cite{Giunti}.}
\label{fig:Sensitivity}
\end{figure}

The 3$\sigma$ and 5$\sigma$ discovery potential for neutrino oscillations is calculated following the prescription in \cite{PDG}. The resulting discovery contours are shown in Figure~\ref{fig:Sensitivity}.  The nominal reactor experiment considered in Table~\ref{tab:DefaultChar} is capable of excluding a large fraction of the  currently preferred parameter space to better than 3$\sigma$ with one year of live-time and at 5$\sigma$ C.L. with 3 years of data taking. Sections~\ref{sec:reac},~\ref{sec:facil}, and~\ref{sec:detect} investigate the impact of the reactor, facility, and detector parameters on the sensitivity of the  experiment. The experimental parameters listed in Table~\ref{tab:DefaultChar} will be used throughout this paper as reference for the nominal experimental arrangement.

\section{Reactor Parameters}
\label{sec:reac}
\subsection{Reactor Power and Operations}

\begin{figure}[htb!]
\centering
  \includegraphics[width=0.5\textwidth]{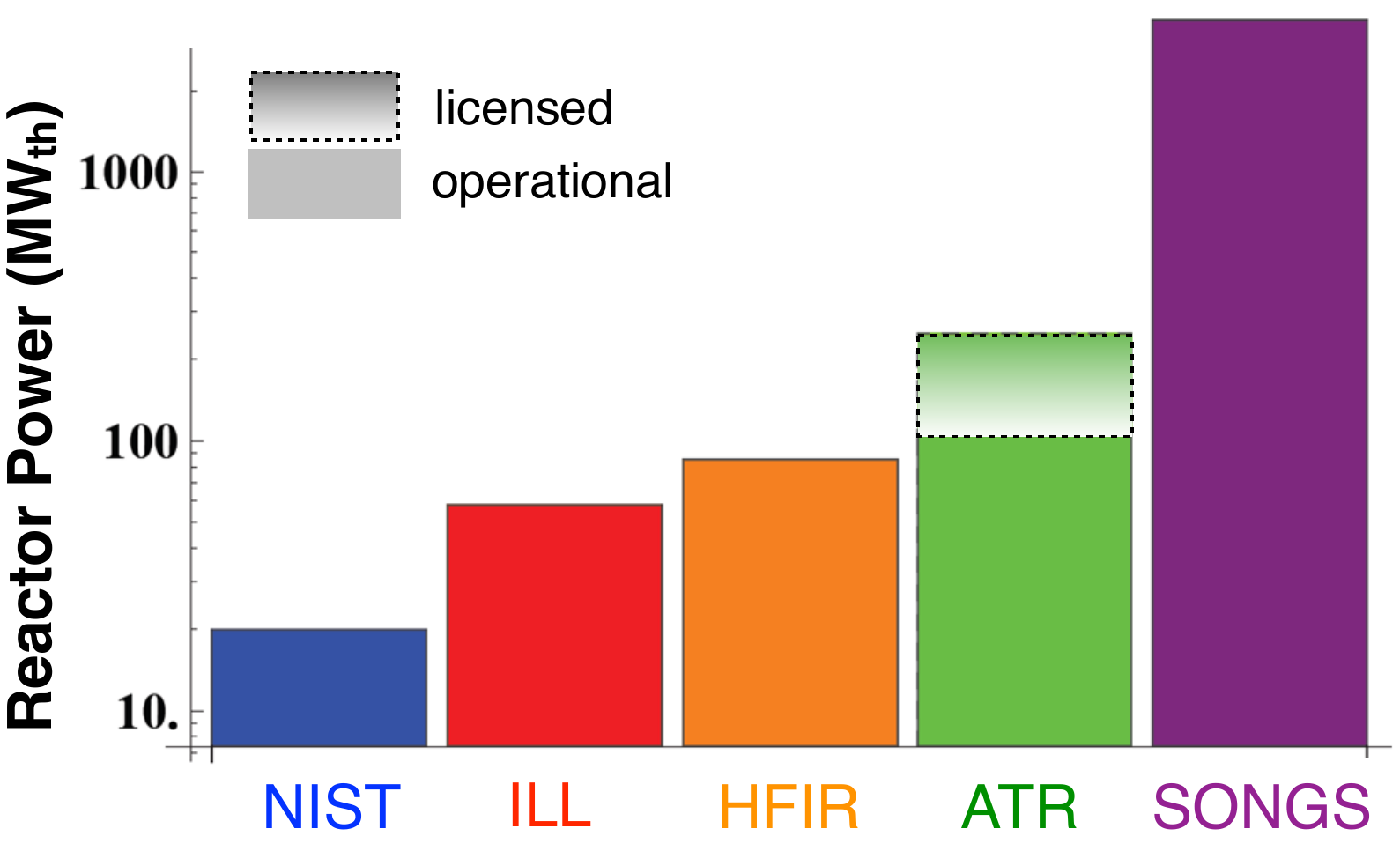}
\hspace{0.1cm}
  \includegraphics[width=0.46\textwidth]{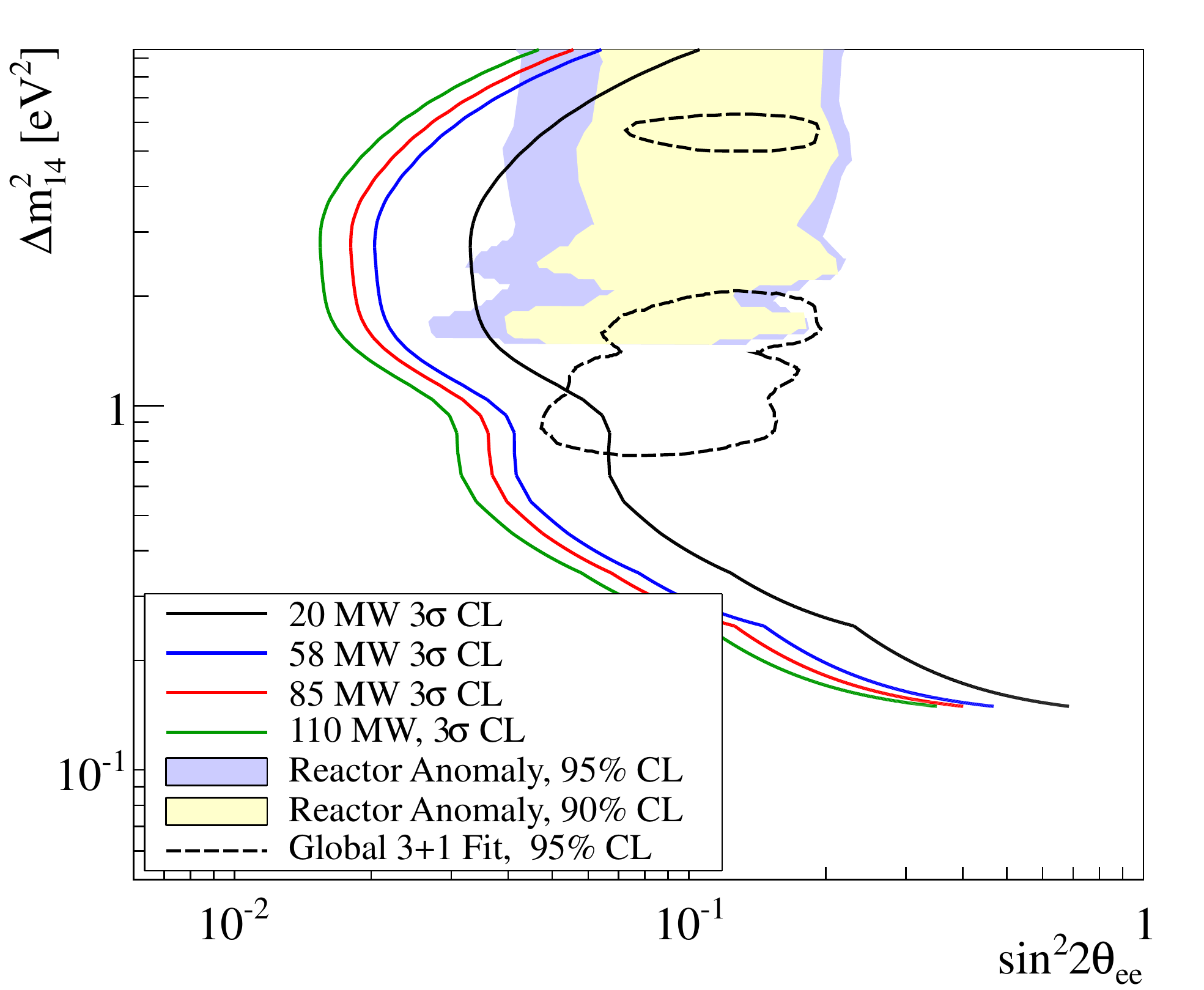}
\caption[]{Left: Comparison of nominal and operational thermal power output for selected reactor facilities. The nominal licensed power output (dashed) is roughly equal to the operational power (solid) for all facilities except the Advanced Test Reactor. Right: Sensitivity of the nominal experimental configuration for various power levels  and 1 year of detector livetime.}
\label{fig:power}
\end{figure}

The flux of  $\nuebar$ emitted from reactors is directly related to its thermal power. The total nominal power capacities for several research and commercial reactor sites are shown in Figure~\ref{fig:power}. At 3\,GW$_{th}$, typical commercial power stations such as the San Onofre Nuclear Power Generating Station (SONGS) \cite{SONGSup, SONGSdown} are about an order of magnitude more powerful than research reactors. Research reactors in the US have a wide range of thermal powers up to a maximum of 250\,MW at the Advanced Test Reactor \cite{ATRschedule}. The variation of the experimental sensitivity of the baseline experimental configuration with the power of several research reactors (or $\nuebar$ flux) is shown in Figure~\ref{fig:power}. The increase in event statistics with thermal power uniformly increases an experiments' sensitivity for all values of $\Delta m^2$.

Nuclear reactors are periodically shut down for refueling and maintenance. As a result, the time-averaged thermal output of a nuclear reactor will be less than its maximum thermal output over that time period because of periods of below-capacity operation and reactor-off periods.  For example,  at ATR the standard operating thermal output of 110~MW is significantly below its maximum licensed thermal output  of 250~MW.  See Figure~\ref{fig:power}. All nuclear facilities undergo reactor-off periods in which reactor refueling and maintenance takes place.  Table~\ref{tab:reactors} summarizes the length and frequency of reactor-off periods at several reactor facilities as well as their reactor-on and -off time.

\begin{table}[!htbp]
  \noindent\makebox[\textwidth]{%
  \begin{tabular}{l|l|l|l|l|l|l} \hline
    Reactor & Power  &  Baselines  & Reactor On  & Reactor Off &  Down-Time & \multirow{2}{*}{Ref.} \\ 
  & (MW$_{th}$) & (m) & (Days) & (Days)  &   \\ \hline
        NIST  & 20 & 4-13  & 42 &10 & $\sim$32\%&  \cite{NISTschedule} \\ \hline
    HFIR & 85 & 6-8  & 24 & 18 &  $\sim$50\% & \cite{HFIRschedule} \\ \hline
        \multirow{2}{*}{ATR}  & 250 (licensed) & 7-8 (restricted)  & \multirow{2}{*}{48-56} & \multirow{2}{*}{14-21} & \multirow{2}{*}{$\sim$27\% }&  \multirow{2}{*}{\cite{ATRschedule}} \\
            & 110 (operational) & 12-20 (full access)  & & & & \\  \hline
    ILL & 58 & 7-9   & 50 & 41 &  $\sim$45\% & \cite{ILLschedule} \\ \hline
    SONGS & 3438 & 24   & 639 & 60 &  8.6\% & \cite{SONGSup, SONGSdown}\\ \hline
  \end{tabular}}
  \caption{\small{Summary of reactor powers, accessible baselines, fuel cycles, and reactor-off times at various reactor facilities.  At ATR, detector access is limited for baselines $<$12\,m. SONGS data is from past operation; this facility is currently shut down. The down-time includes estimates for seasonal shutdowns and maintenance periods.}}
  \label{tab:reactors}
\end{table}


Reactor-off time and operation at reduced power directly reduce the total annual $\nuebar$ event statistics yielding the same net effect as a lower thermal power capacity. The experiment's sensitivity decreases with lower power as illustrated in  Figure~\ref{fig:power}. However, reactor-off time provides an opportunity to measure the rate and energy distribution of backgrounds.  A detailed understanding of the spatial and energy distribution of backgrounds is critical for a precision experiment at short distances from the reactor.  This topic will be discussed in greater detail in Section~\ref{sec:Bkgs}. The optimum ratio of background measurement time relative to $\antinue$ signal time will depend on the total signal statistics, the dominant uncertainties, as well as the signal-to-background ratio. This optimization should be carried out as part of a detailed design process. For the research reactors in Table~\ref{tab:reactors} the $\antinue$ signal time will be between 50-70\% of calendar time. 

\subsection{Reactor Fuel}

Commercial nuclear power stations use conventional nuclear fuel comprised of a mixture of U and Pu isotopes while some research facilities operate with highly-enriched uranium (HEU). The four isotopes $^{235}$U, $^{238}$U, $^{239}$Pu, and $^{241}$Pu produce $>$99.9\% of all \antinue produced in a reactor. In HEU cores nearly all fissions are accounted for by $^{235}$U.  Table~\ref{tab:fissionfractions} gives the main fission isotopes in the two reactor fuels and their relative contributions to the total fission rate. 

\begin{table}[htbp]
  \centering
  \begin{tabular}{c| c| c} 
\hline
    \multirow{2}{*}{Fuel Isotope} & \multicolumn{2}{|c}{Time-Averaged Fission Fraction}\\ \cline{2-3} 
    & Conventional Fuel & HEU fuel\\ 
    \hline
    $^{235}$U & 0.59 & $>$0.99 \\
    $^{238}$U & 0.07 & $<$0.01 \\
    $^{239}$Pu & 0.29 & $<$0.01 \\
    $^{241}$Pu & 0.05 & $<$0.01 \\
    \hline
      \end{tabular}
  \caption{Approximate time-averaged fuel compositions for various reactor cores.  Fractions for conventional ~\cite{Kopeikin} and HEU reactors~\cite{Glaser}, respectively.}
  \label{tab:fissionfractions}
\end{table}

\begin{figure}[htb!]
\centering
    \includegraphics[trim=4cm 3.35cm 2cm 1cm, clip=true, width=0.44\textwidth]{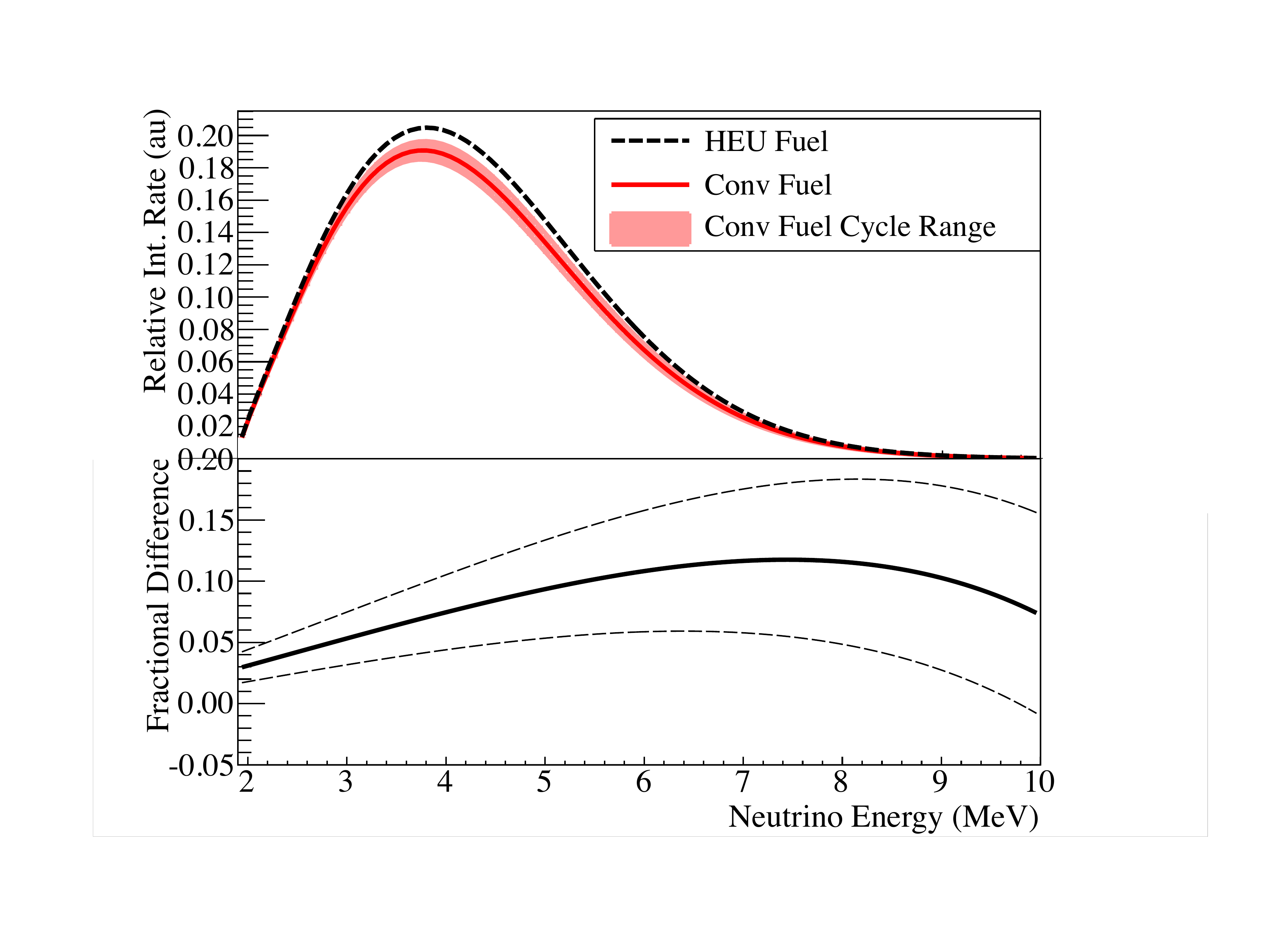}
    \includegraphics[trim=0.1cm 0.5cm 0.1cm 0.1cm, clip=true, width=0.48\textwidth]{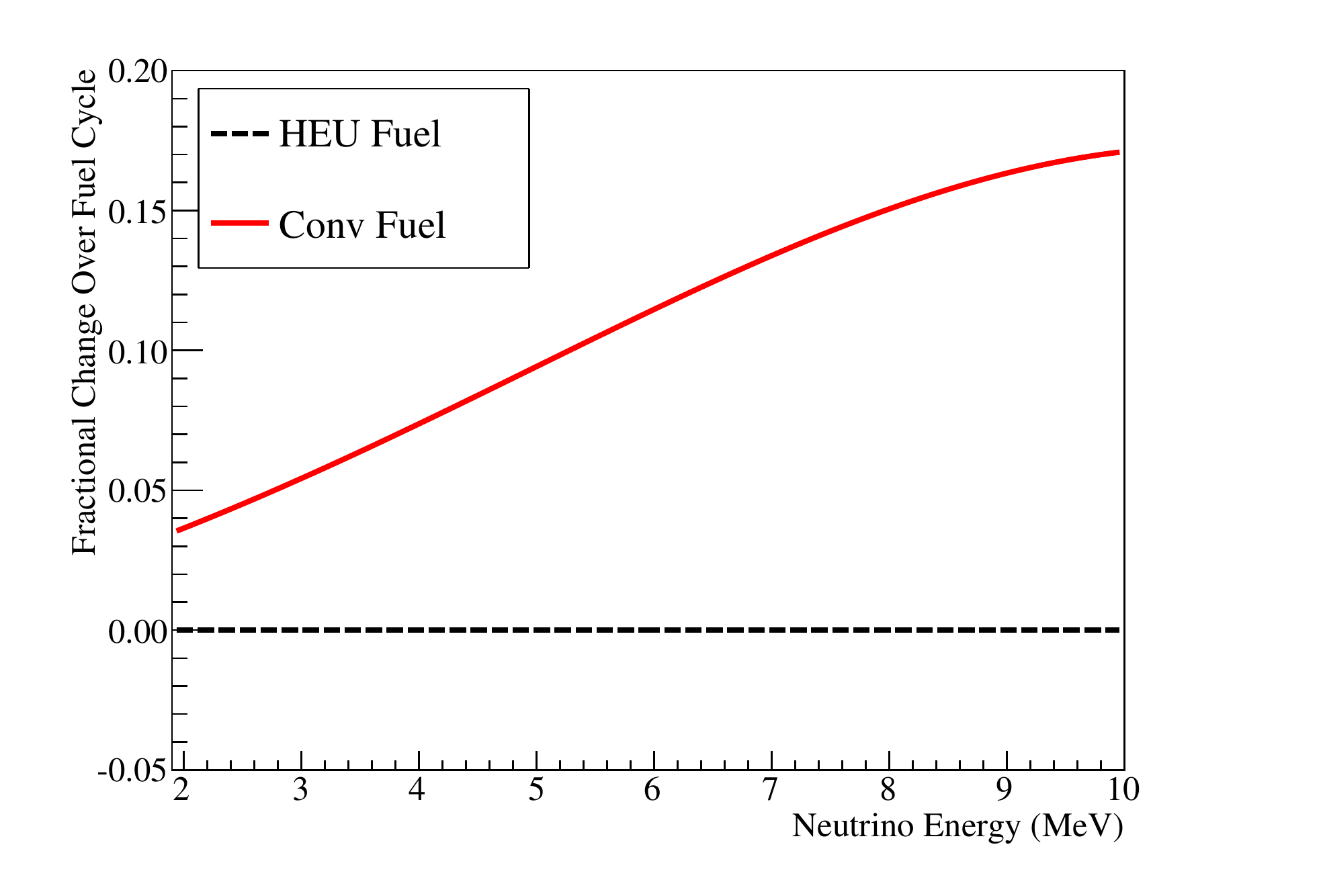}
\caption{Left: Detected reactor $\nuebar$ spectra from conventional commercial reactor and HEU reactor fuel (top panel), and their spectral differences (bottom panel). The band indicates the change in the spectral shape over one fuel cycle. Right: Fractional change in spectrum over one fuel cycle for conventional and HEU reactors.}
\label{fig:FuelSpectra}
\end{figure}

\begin{figure}[htb!]
\centering
\includegraphics[width=0.45\textwidth]{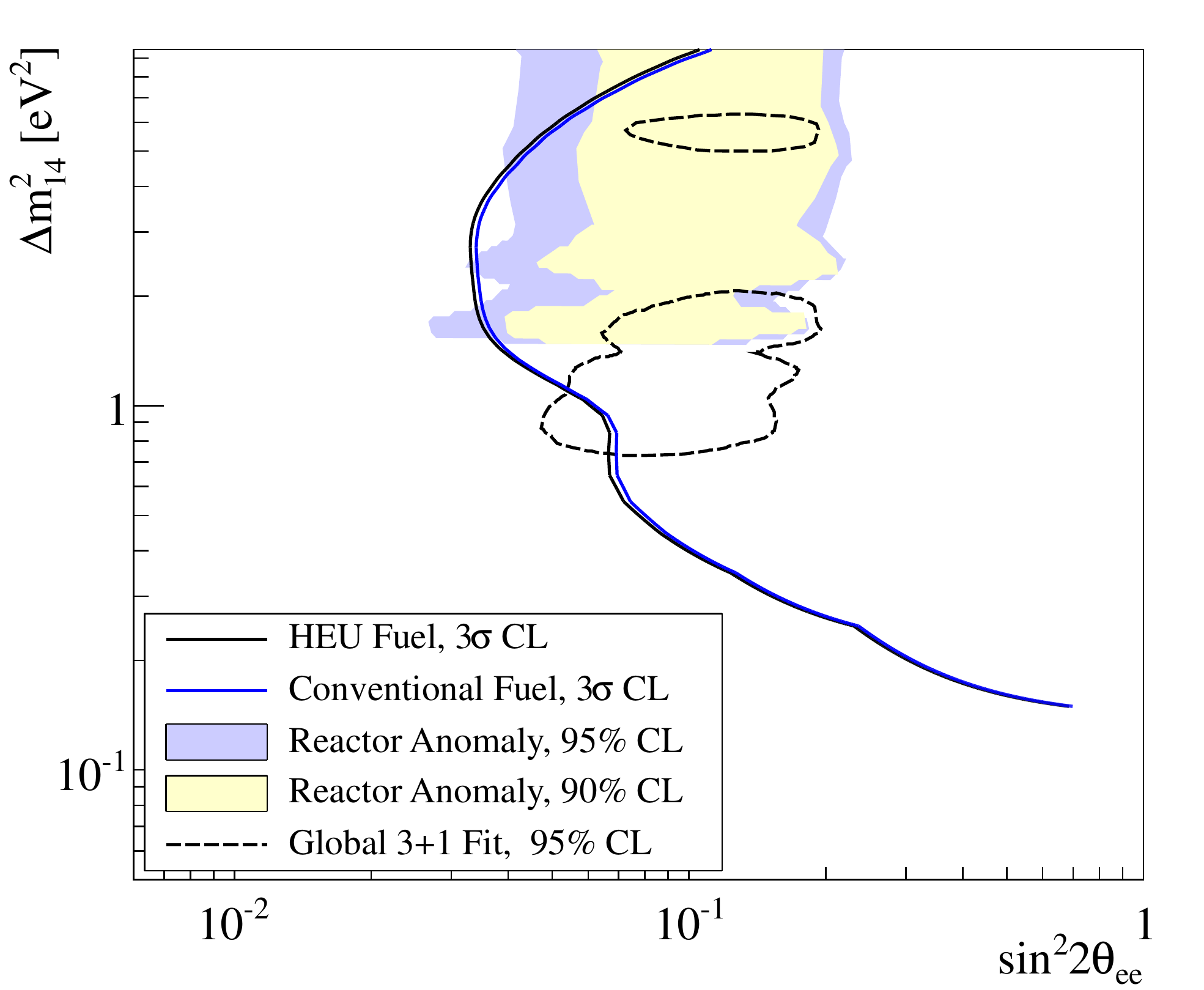}
\caption{Sensitivities of the nominal reactor experiment for conventional and HEU fuel reactor cores. The use of HEU fuel leads to minor improvements in the overall sensitivity.}
\label{fig:FuelSensitivity}
\end{figure}

The $\nuebar$ spectrum and rate per fission is different for each of these isotopes~\cite{Vogel} and the flux and spectrum of reactor $\antinue$ are the sum for all isotopes in the reactor core. The time-averaged detected $\nuebar$ spectra from HEU and conventional fuels are compared in Figure~\ref{fig:FuelSpectra}.  
The integrated flux differs by roughly 8\%, and the time-averaged spectral differences are 10\% or less. 
The time evolution of the isotopic fuel composition creates a time-dependent spectral shape. For the spectral range shown in Figure~\ref{fig:FuelSpectra} the typical fission fractions for the average fuel were evaluated and compared to the upper and lower fractions at the beginning and end of fuel cycle \cite{Kopeikin}.  Using the spectral shapes from \cite{Vogel} together with the isotope fraction the combined spectrum can be calculated. When convoluted with the standard IBD cross section we obtain the detected $\antinue$ spectrum. The differences in the energy spectrum only have a small impact on the experiment's sensitivity to short-baseline neutrino oscillation as shown in Figure~\ref{fig:FuelSensitivity}.

In addition to differences in the integrated flux and spectral shape, the uncertainties spectral shape vary between the dominate fission isotopes.  These arise from a combination of the statistical and measurement uncertainties of fission isotope beta spectra \cite{ILL} and uncertainties in the conversion of electron spectra to corresponding $\nuebar$ spectra~\cite{Schrek1, Schrek2, HuberAnomaly}.  Spectral uncertainties for $^{235}$U range from 1.8-3.2\%  in the range from 2-6\,MeV, while for $^{239}$Pu they increase to 1.9\%-5.7\% and 2.5\%-5.0\% for $^{241}$Pu~\cite{Christine}.  Because of these considerations, spectral uncertainties will be lower for HEU fuel than for conventional fuel.  The change in sensitivity resulting from this difference in spectral uncertainty for the two fuel types is shown in Figure~\ref{fig:FuelSensitivity}.  HEU fuel provides a minor improvement in sensitivity as a result of the lowered spectral uncertainties.  The benefits of lower spectral uncertainties will be amplified in the case of detectors with limited baseline ranges or position resolution, as will be discussed in Section~\ref{subsec:Versus}.

\subsection{Reactor Core Size and Dimensions}

\begin{figure}[htbp]
   \centering
   \includegraphics[width=0.48\textwidth]{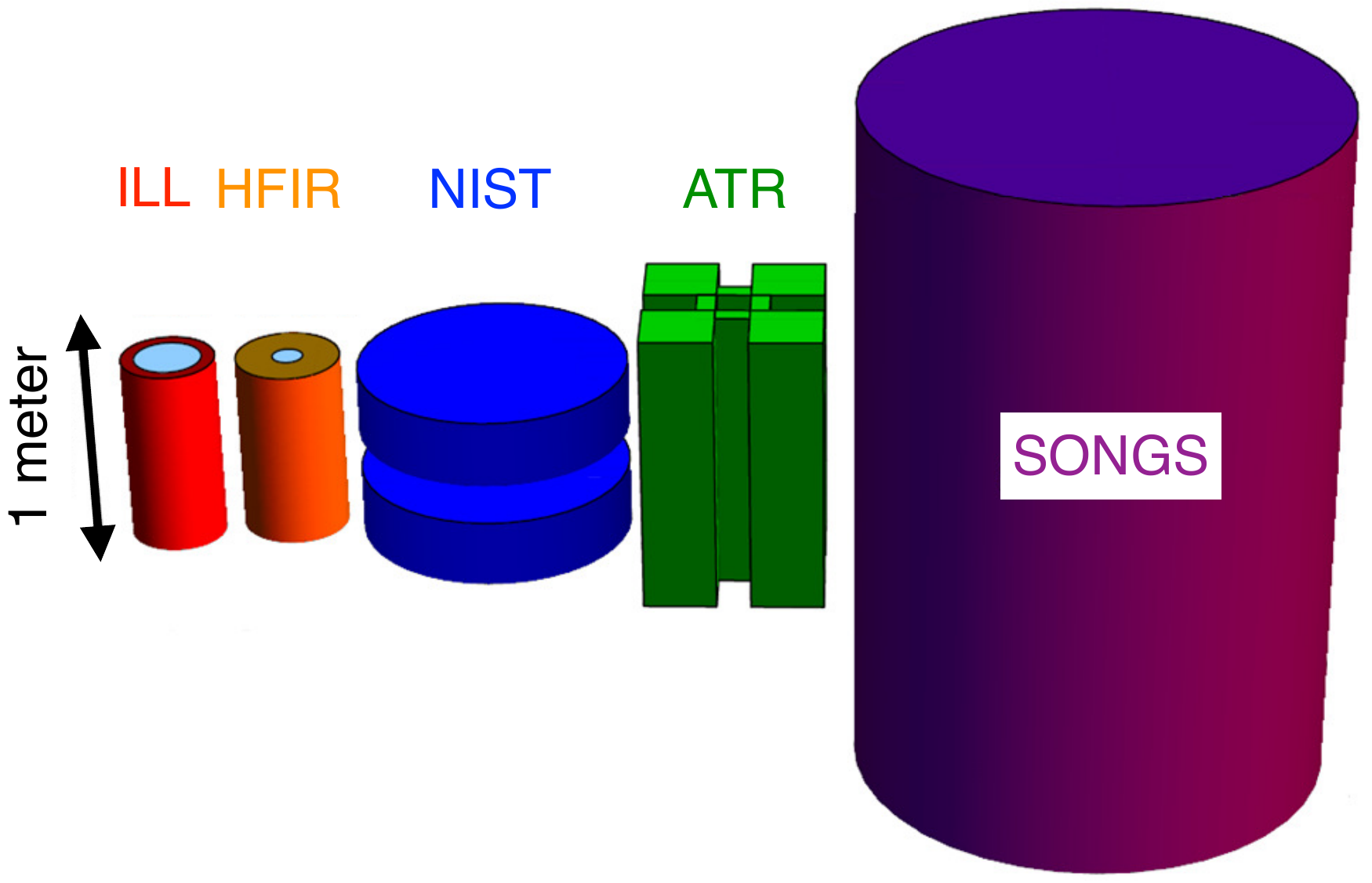}
   \includegraphics[width=0.48\textwidth]{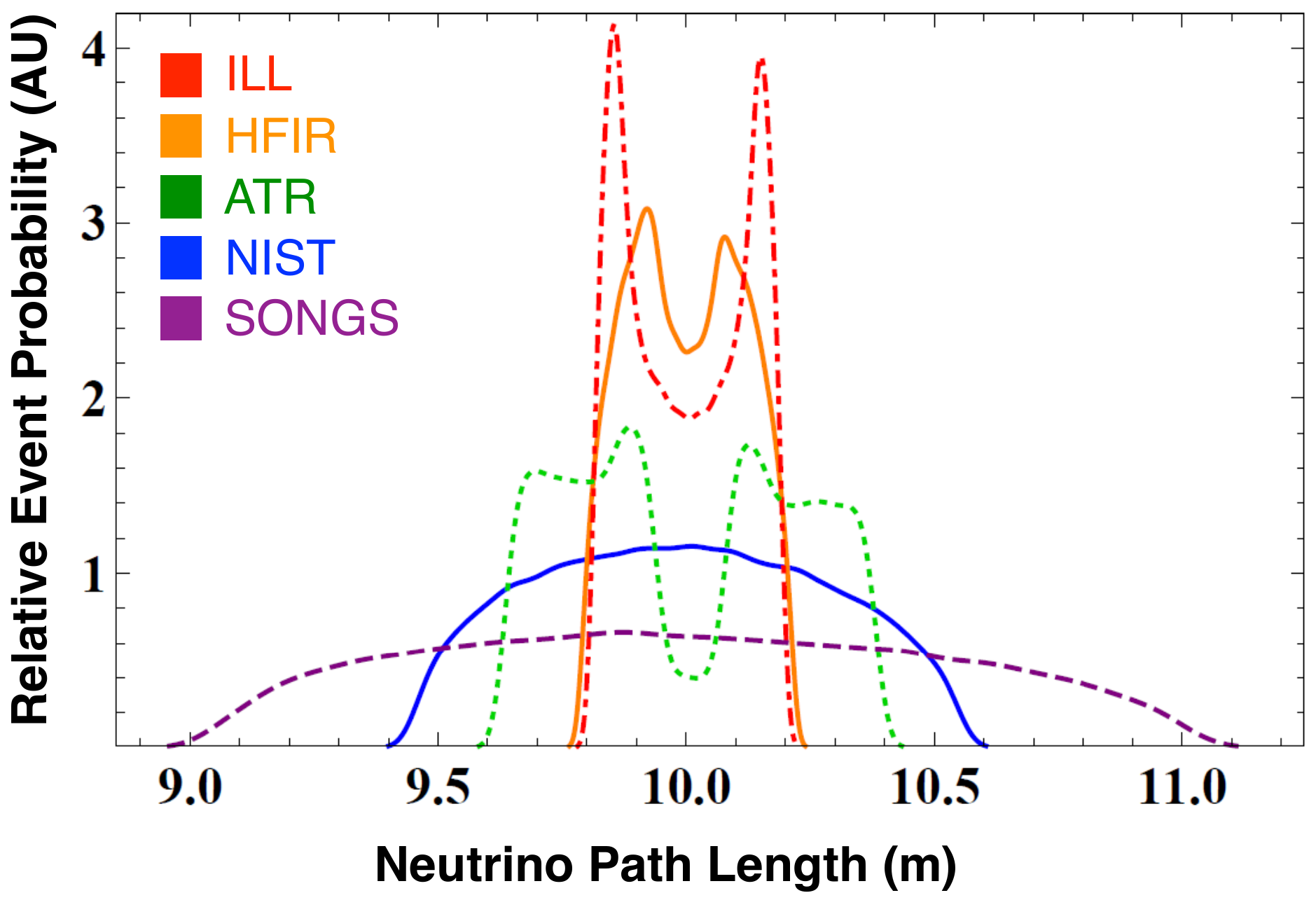}
   \caption{Left: Approximate size and shape of fuel distribution in five reactors including the US research reactors ATR, NIST, and HFIR.  SONGS is  a conventional power reactor, for comparison. Right: Relative event probability versus neutrino path lengths for several finite-sized reactor cores and a point-like detector horizontally displaced from the vertical midpoint of each reactor at a distance of $r=10$\,m.}
   \label{fig:Core}
\end{figure}

Reactor cores come in a wide variety of shapes and sizes. The $\antinue$ production in a reactor follows closely the distribution of the fuel assemblies. For the purpose of a neutrino oscillation experiment at very short baselines the main difference  between commercial power reactors and research reactors is in their size.  Conventional pressurized water power reactors are typically 3-4\, m in height and diameter, while research reactors tend to be smaller in size with dimensions as compact as $\sim$0.5\,m or less.   Figure~\ref{fig:Core} provides an illustration of the variety of shapes and sizes of various research and conventional reactor cores.

For neutrino oscillation searches it is required that the size of the reactor core and thus the spread in neutrino path lengths to be less than the oscillation wavelength to avoid a wash-out of the oscillation signal. 
Figure~\ref{fig:Core} shows the spread in path lengths between the finite-sized reactor cores and a point-like detector  horizontally displaced from the vertical midpoint of each reactor at a distance of $r=10$\,m. The distance, $r$,  is defined between the geometric center of the core and the point-like detector.  A distance of $\mathcal{O}$(10\,m) represents a typical distance for very short baseline reactor experiment.  We assume that neutrinos are produced and emitted uniformly throughout the core region.  We define the path length, $l$,  as the distance between the points of $\antinue$ production and observation. Convolving the path length distribution with 1/$R^2$, determines the expected spatial distribution for the relative probability of $\antinue$ interactions in the detector. We note that a detector  with finite position resolution or segmentation adds additional smearing to the observed path length distribution and the observed oscillation effect.  


For finite-sized reactor cores of any geometry we can calculate the average path length, $\overline{l}$ and the RMS of the path lengths as

\begin{eqnarray}
\overline{l}=\frac{1}{volume}\int_{vol}\sqrt{x^2+y^2+z^2}dxdydz \\
{l_{rms}}=\sqrt{\frac{1}{volume}\int_{vol}\sqrt{x^2+y^2+z^2-\overline{l}^2}dxdydz}
\end{eqnarray}

\begin{figure}[htbp]
   \centering
  \includegraphics[trim=2cm 14cm 18cm 1.5cm, clip=true, width=0.44\textwidth]{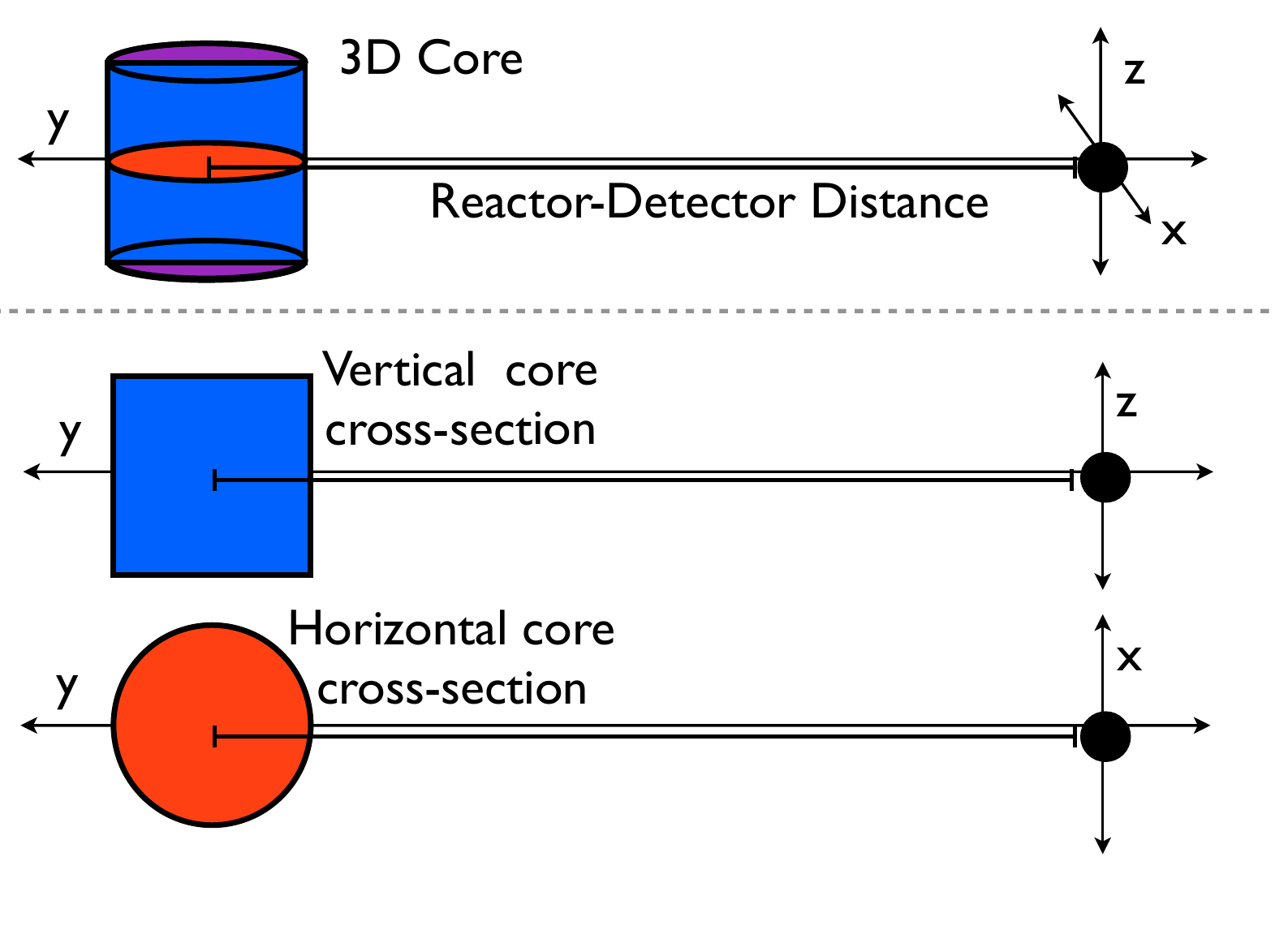}
\hspace{0.1cm}
  \includegraphics[width=0.48\textwidth]{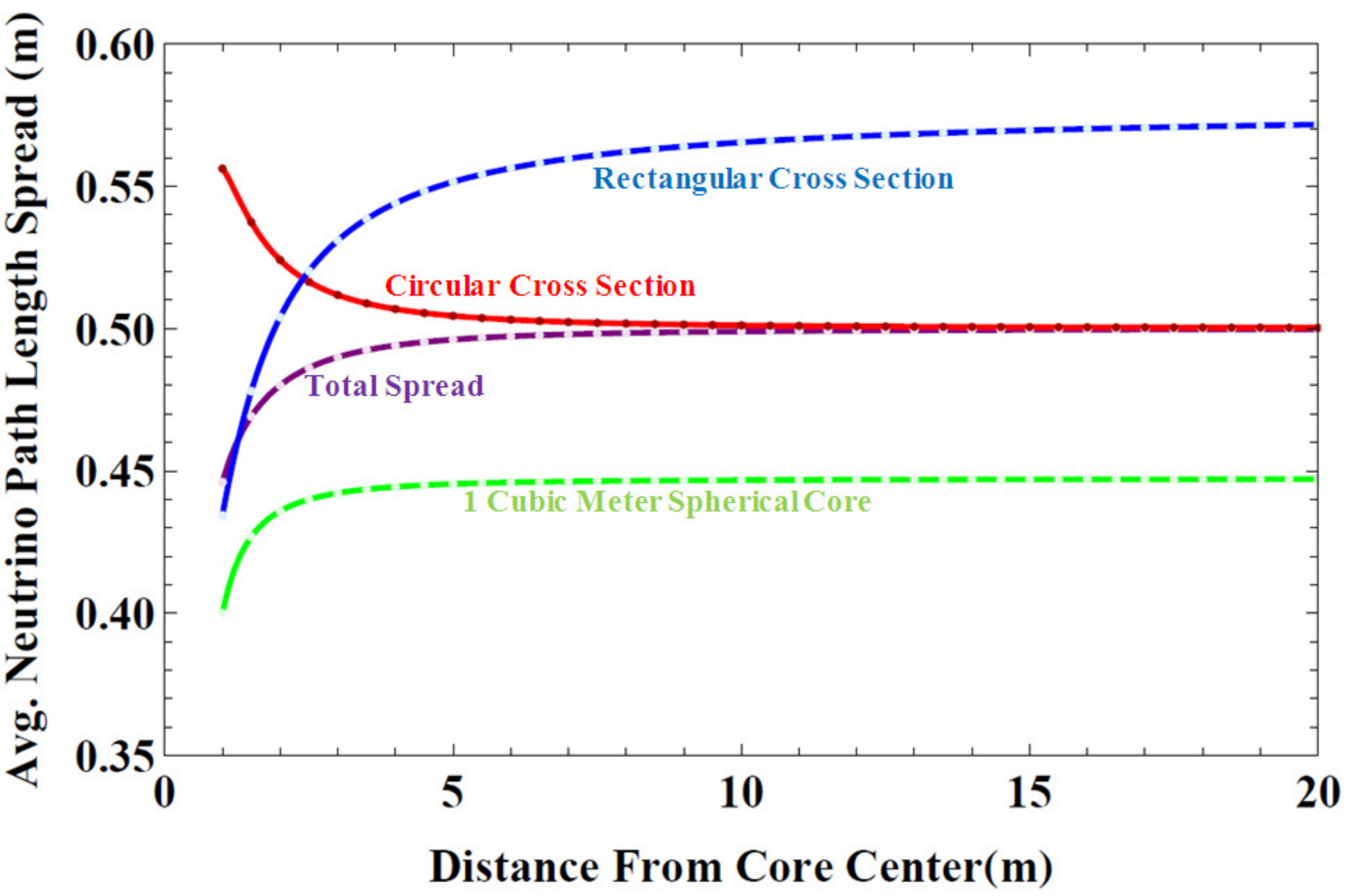}
   \caption{Left: Geometries, cross-sections, and path lengths for a cylindrical reactor core and point-like detector. Right: Corresponding path length spreads for a 1-m cylindrical core. The differences in path length spreads contributed by the circular horizontal dimensions and the rectangular vertical dimensions of the core are a small correction on top of the overall magnitude of the spread, which is defined by the overall core size.}
   \label{fig:CrossSections}
\end{figure}

\begin{figure}[htbp]
   \centering
  \includegraphics[width=0.46\textwidth]{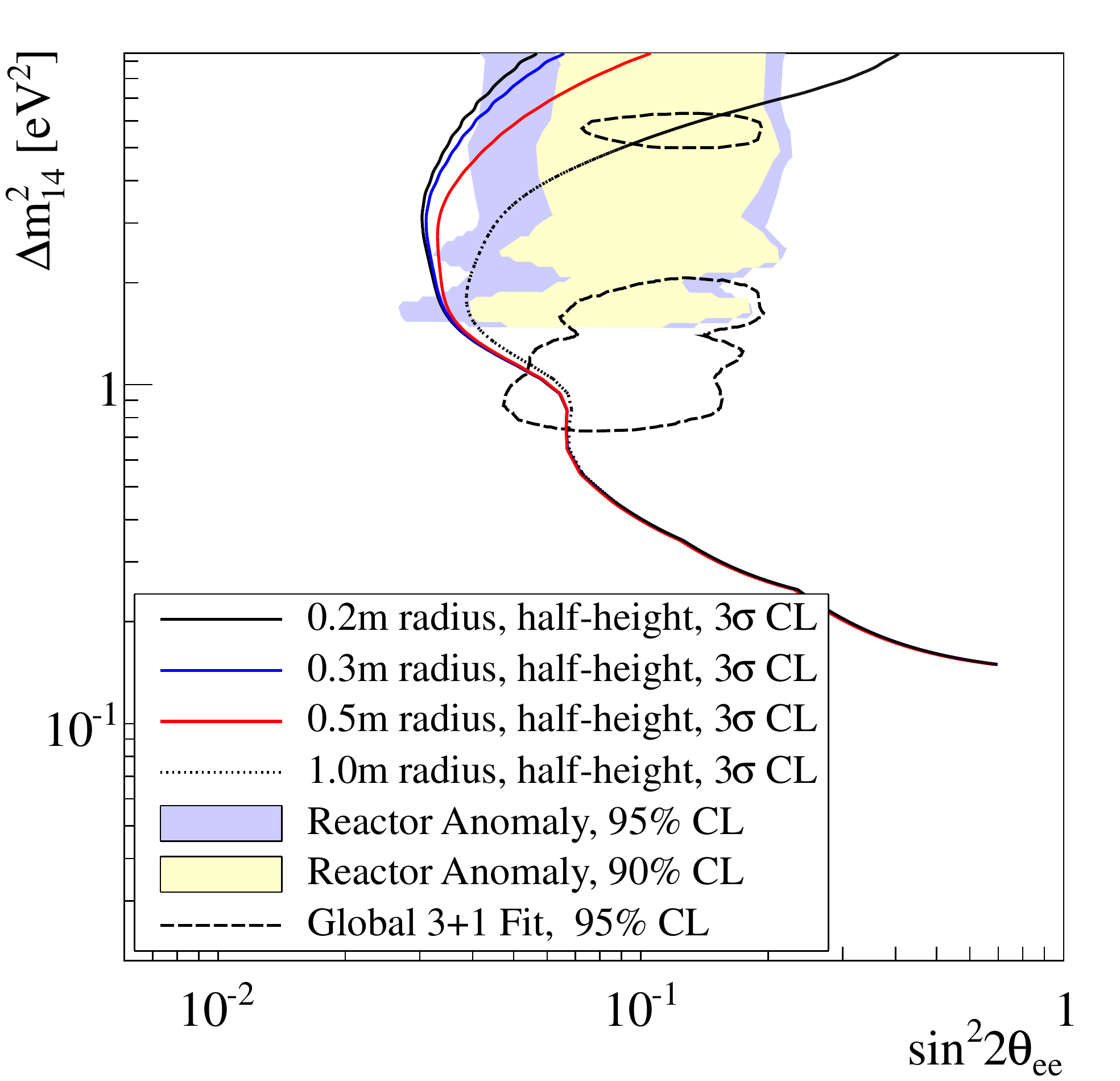}
\hspace{0.1cm}
  \includegraphics[width=0.46\textwidth]{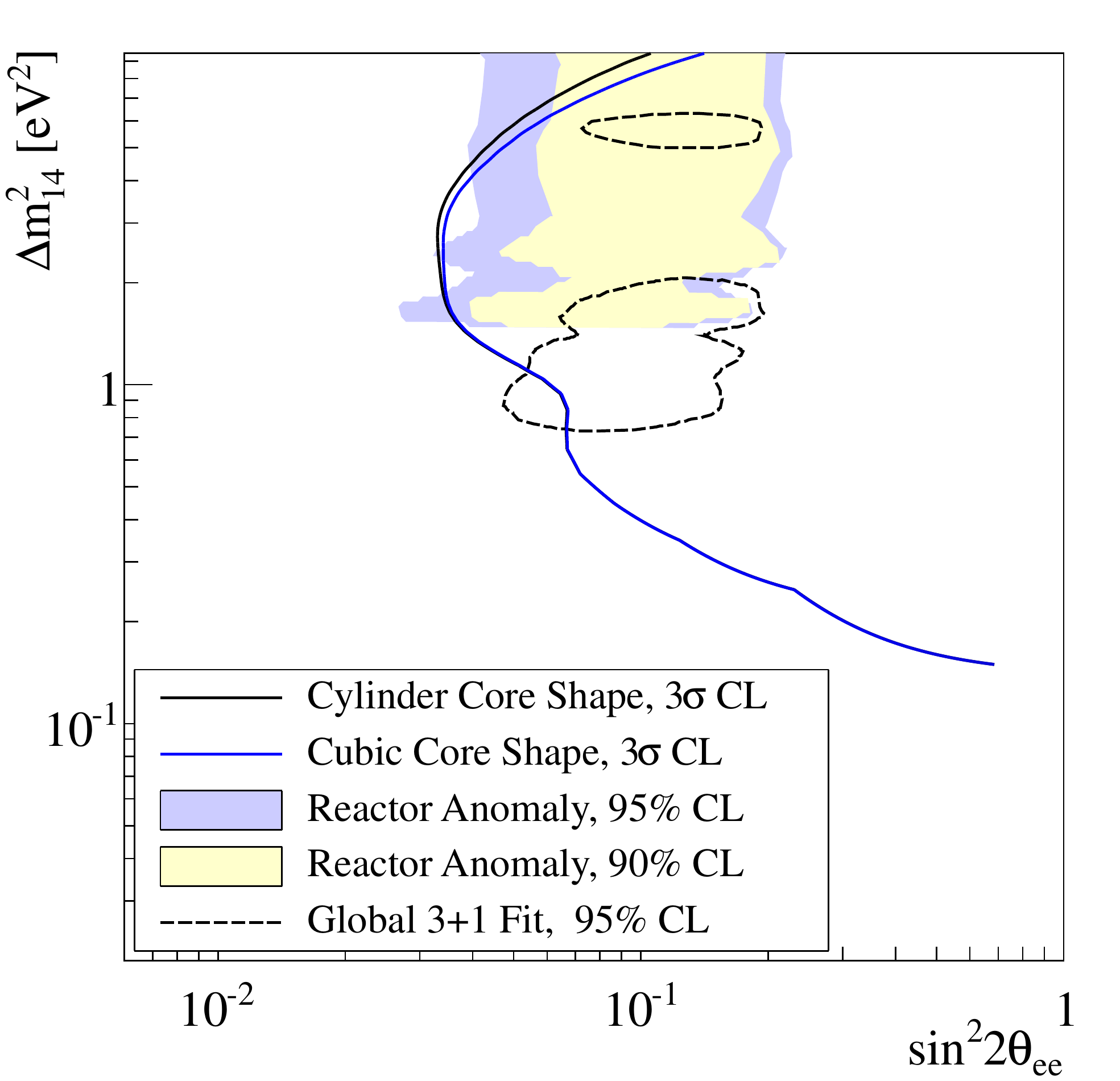}
   \caption{Variations in the discovery potential of the nominal reactor experiment with a half-height core of various radii (left) and core geometries (right).  The overall spread in neutrino path lengths is dominated by the overall dimensions of the core.}
   \label{fig:ShapeSize}
\end{figure}

Figure~\ref{fig:CrossSections} shows the average neutrino path length spread for different  core geometries as a function of distance from the reactor. At $d>5$\,m the RMS spread in path lengths approaches $\sim$0.5-0.6\,m for a reactor of 1\,m height and diameter. The variation due to the core shape is significantly smaller than the total magnitude of spread. For all practical considerations and for realistic distances of $d>4$\,m from the reactor the shape of the core and fuel distribution is only of secondary consideration. 
For highly asymmetric cores such as long, cylindrical arrangements choosing the orientation of the detector with respect to the symmetry axis of the core can be used to reduce the spread in neutrino path lengths.   Figure~\ref{fig:ShapeSize} shows the discovery potential for the nominal reactor experiment for varying dimensions of the reactor core.   The sensitivity to higher $\Delta m^2$ values is lost as the core width is increased.  The overall core dimensions dominates the spread in neutrino path lengths  while the reactor shape only plays a secondary role.  

\section{Facility Parameters}
\label{sec:facil}

\subsection{Experimental Area and Detector Volume}
\label{subsec:Lengths}

A key signature of neutrino oscillation is the distortion of the measured spectrum with energy and the variation of the energy spectrum with baseline.   The baseline dependence of the energy spectrum can only be observed in a detector whose active length comprises more than a small portion of one oscillation period. Ideally, one would map out the oscillation over an entire wavelength or more. The availability of experimental space in a reactor facility limits the accessible baselines as well as the total active detector volume. The accessible baselines listed in Table~\ref{tab:reactors} provide a distance range of up to 9\,m at one particular facility. This allows the placement of point-like detectors in various locations, or  the construction of an extended detector in the radial direction. For any significantly extended detector the change in event rate due to the 1/R$^2$ law has to be taken into account. 

\begin{figure}[htbp]
   \centering
  \includegraphics[width=0.46\textwidth]{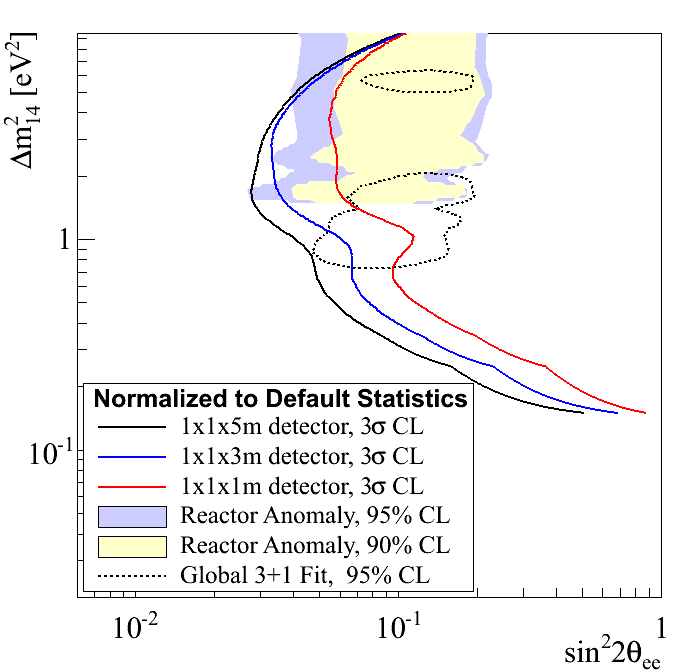}
\hspace{0.1cm}
  \includegraphics[width=0.46\textwidth]{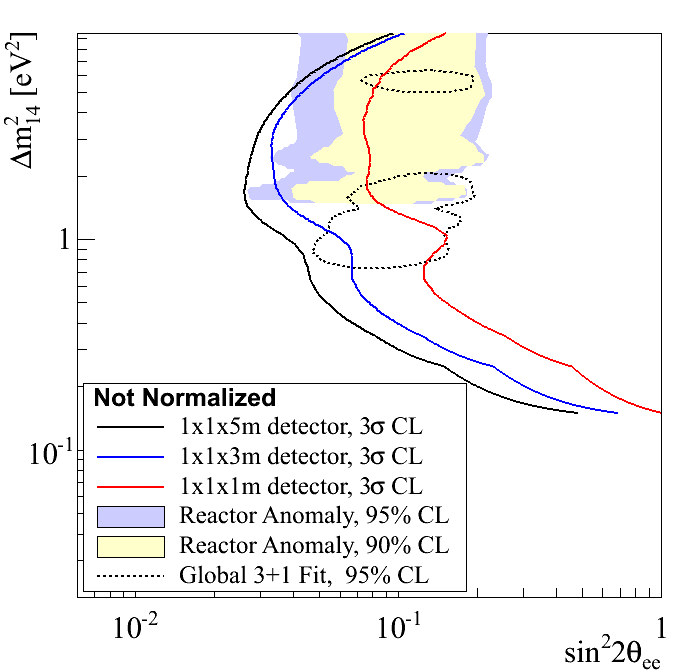}
   \caption{Variation in sensitivity of the default experiment with detector length.  Left: Event statistics is normalized to the default experiment in each case.  Both the maximum achievable sensitivity and the range of accessible $\Delta m^2$ are increased with increased detector length. Right:  Un-normalized case demonstrates the additional effect of increased statistics with larger detector volume.}
   \label{fig:DetWidth}
\end{figure}

An increase in detector length has two effects: It increases the total overall event statistics of the experiment and it samples a larger fraction of the possible oscillation wavelength. Both the maximum achievable sensitivity and the overall range of accessible $\Delta m^2$ values improve with an increase in detector length.  In particular, sensitivity to lower values of $\Delta m^2$ is improved with a longer detector as oscillations with longer wavelength are sampled.  Figure~\ref{fig:DetWidth} shows the change in sensitivity of an experiment with the default characteristics for three different detectors of 1, 3, and 5\,m lengths.  Figure~\ref{fig:DetWidth} shows on one side the overall sensitivity of the three detectors with the different event statistics due the various detector lengths and on the other hand the same experimental arrangement normalized to the statistics of the default experiment. The normalized case illustrates the impact of the detector length or baselines while the un-normalized situation highlights the decrease in event rate with 1/$R^2$. 

The detector cross-section is typically limited by space constraints inside the containment building near the reactor core.  As the overall target mass and thus statistics at each baseline scale with the cross-sectional area of the detector, one can consider scaling the cross-sectional area of the detector as a function of baseline to counteract the effect of the 1/$R^2$ reduction in flux.

\subsection{Reactor-Detector Distance}
\label{subsec:Distance}

For extended core and detector geometries, the distance between the reactor and the detector is not uniquely defined. 
We therefore choose the reactor-to-detector distance, $r$, as the distance between the center of the reactor core and the closest point to the reactor in the active detector region while the detector length, $d$, describes the radial length of the active detector volume.  The reactor-to-detector distance, $r$, at a minimum, is comprised of the extent of the reactor itself and the thickness of the containment and shielding structures surrounding the core. In practice some passive shielding will be required to operate a $\antinue$ detector in the vicinity of a reactor. 
The accessible baselines for various reactor facilities listed in Table~\ref{tab:reactors} take 0.5~m additional space into consideration for this purpose.  The closest accessible baseline is site specific and referred to as $r_{min}$.  As is evident in Table~\ref{tab:reactors}, research reactors provide an opportunity for improved measurements of the reactor $\antinue$ flux and spectrum at the shortest baselines to date. Facilities such as NIST may provide access to baselines as short as 4\,m or less.

\begin{figure}[htbp]
   \centering
  \includegraphics[width=0.52\textwidth]{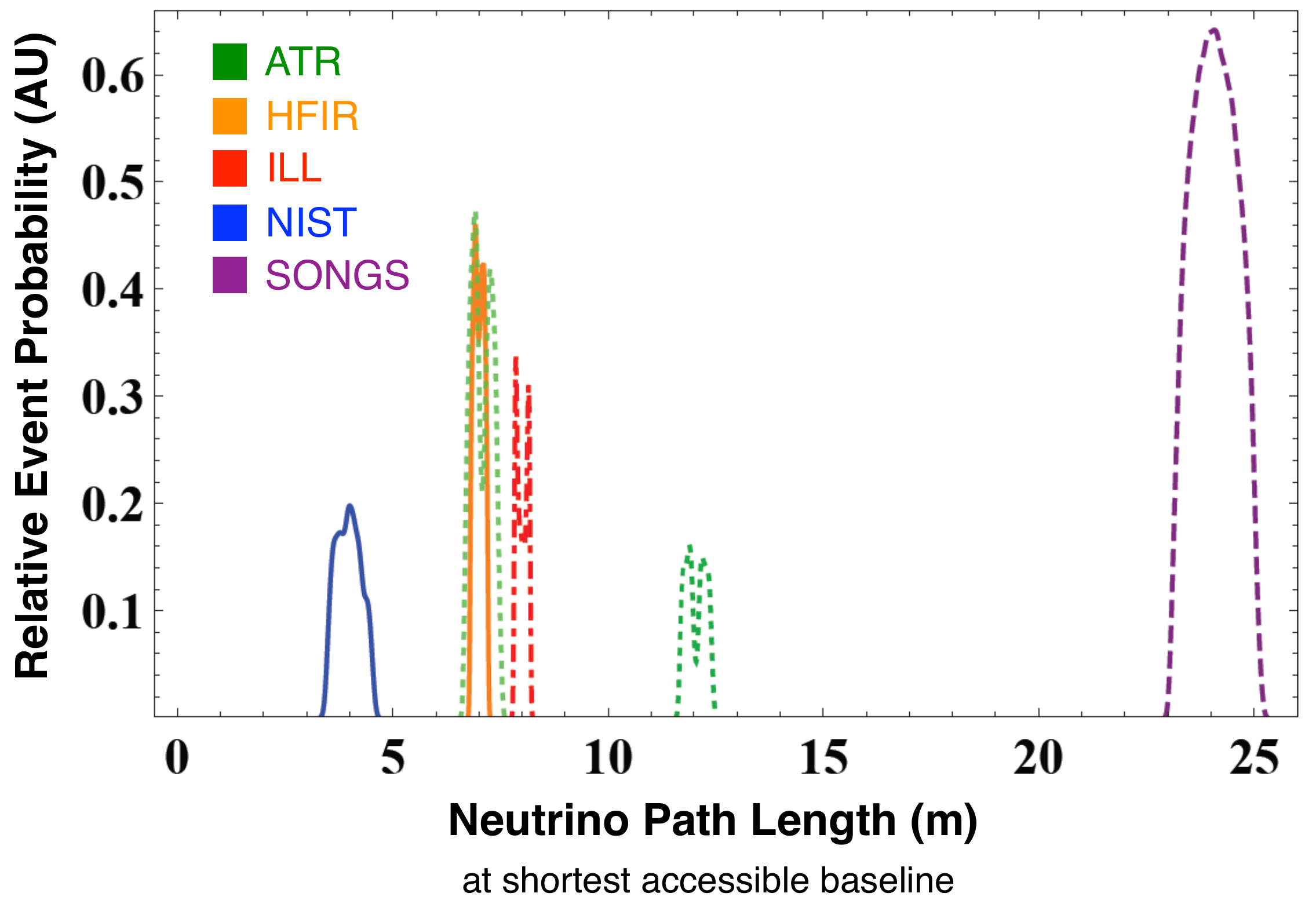}
  \includegraphics[width=0.45\textwidth]{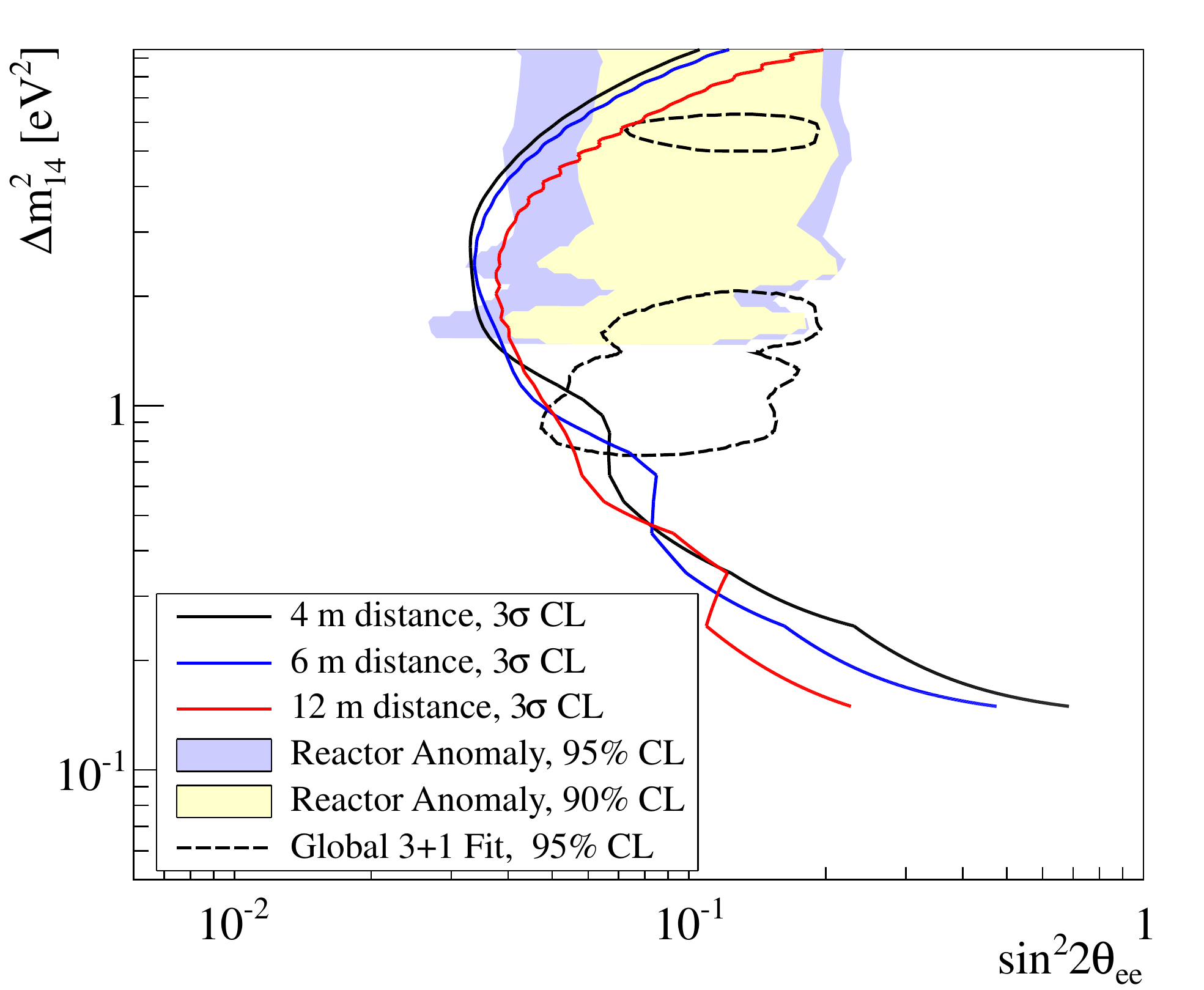}
   \caption{Left: Relative probability of $\nuebar$ interactions in a point-like detector at the closest possible distance from the respective reactor cores. The spread in neutrino path lengths reflects the size and shape of the reactor cores.  The peak areas give the relative reactor power scaled by 1/$R^2$.  Right: Sensitivity of the default configuration for closest accessible baselines of 4, 6, and 12\,m respectively.  To emphasize only the effect of baseline, the normalization is adjusted to provide equal total event statistics.}
   \label{fig:Distance}
\end{figure}

In maximizing event statistics and experimental sensitivity, the design of an experiment is a trade-off between distance, spread in neutrino path length, and relative reactor power. Figure~\ref{fig:Distance} shows the spread in neutrino path lengths from finite-sized reactor cores as seen by a point-like detector at the closest accessible baseline $r_{min}$. The spread in neutrino path lengths reflects the size of the reactor cores while the area of each distribution indicates the relative reactor power scaled by 1/$R^2$. 

The right panel of Figure~\ref{fig:Distance} shows the sensitivity of the default experiment for $r_{min}=$ 4, 6, and 12\,m respectively.  To emphasize the effect of varying $r_{min}$, the normalization is adjusted to maintain the same total statistical sensitivity in each case.  The closest accessible baseline, $r_{min}$, impacts  the high-$\Delta m^2$ sensitivity of the experiment.   This is due to the fact that finite energy resolution tends to wash out the observed oscillation with distance from the reactor core. 

\begin{figure}[htbp]
   \centering
      \includegraphics[trim=0.5cm 11cm 9cm 2cm, clip=true, width=0.7\textwidth]{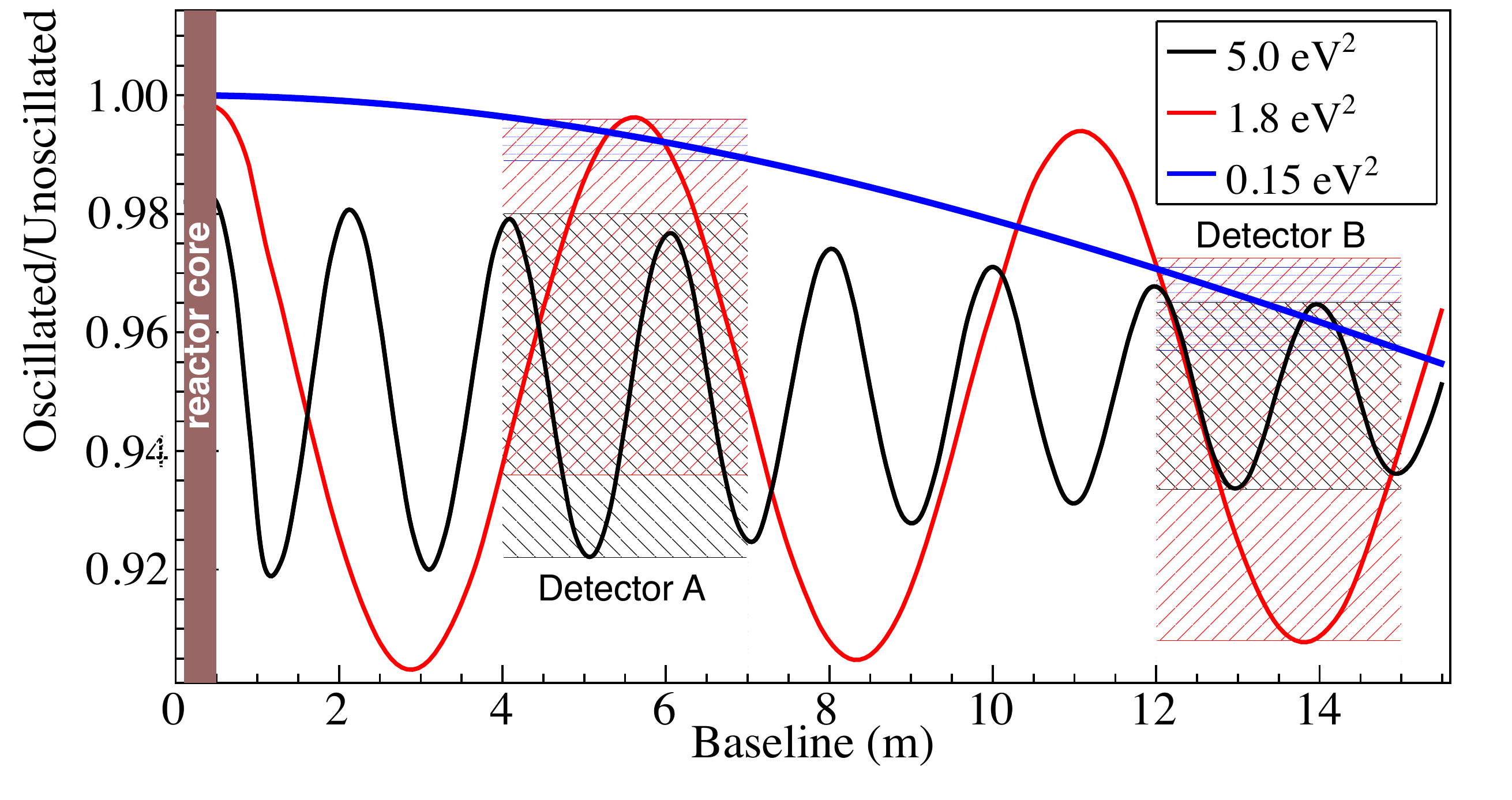}
   \caption{Oscillation of 4\,MeV $\nuebar$ for $\Delta m^2$ of 6.0, 1.8, and 0.15 eV$^2$ as a function of baseline. The shaded regions indicate the position and radial dimension of the detectors for the default experimental configuration. The dimensions of the reactor core are shown in solid color. 
     Oscillation curves include position smearing resulting from position resolution and from the finite reactor core size.  Boxed areas indicate the range of oscillated/unoscillated oscillation amplitudes covered by each case.
For the large $\Delta m^2$ value (5 eV$^2$), the 3\,m detector length encapsulates more than a full period, and the damping of the oscillation amplitude due to finite detector resolution reduces the visible oscillation at farther distances.  For medium $\Delta m^2$ values (1.8 eV$^2$), the 3\,m detector length allows for significant observation of oscillation a both distances.  For very small $\Delta m^2$ (0.15 eV$^2$), the variation in oscillation from detector font to back is larger at 12\,m than at 4\,m, although both differences are small, as the 3\,m detector length encapsulates only a small portion of the total period.  Two detectors at different baselines may be needed to probe regions of small $\Delta m^2$.}
     \label{fig:Match}
\end{figure}

The effect of $r_{min}$ on the experiment's sensitivity to various ranges of $\Delta m^2$ is dependent on the total detector length $d$.  Longer detectors help improve the sensitivity to lower values of $\Delta m^2$ but do not compensate completely in case of larger $r_{min}$.  This is illustrated in Figure~\ref{fig:Match} for a 3\,m-long detector placed at 4\,m and 12\,m respectively from the reactor core. Figure~\ref{fig:Match} illustrates how a detector of chosen length samples different fractions of the oscillation periods and oscillation amplitudes for various $\Delta m^2$.  For larger $\Delta m^2$, the best achievable sensitivity at large $r_{min}$ is lower even with longer detectors because the amplitude of the oscillation effect is diminished.  For low $\Delta m^2$, only a small portion of the oscillation period fits inside the detector length at both distances, indicating the need to sample a much longer range of baselines.  For the favored mass splitting the oscillation period is sampled well at both chosen baselines.



If the oscillation length were known {\em a priori},  a detector of suitable length could be placed between the first oscillation minimum and subsequent maximum to maximize the observed oscillation difference or around the oscillation minimum to observe the turning point. Ideally, such a detector would be movable to measure both the difference between the oscillation maximum and minimum and the turning point around one of the oscillation extreme. A movable, extended detector can also help mitigate possible backgrounds or systematic effects that can mimic this oscillatory signature.  In the case of longer wavelengths and given the facility constraints at research reactors, multiple, radially extended detectors may be necessary to allow for a comprehensive search and discovery of neutrino oscillations with unknown $\Delta m^2$. 



\newpage

\section{Backgrounds}
\label{sec:Bkgs}

The overall magnitude and spectral shape of backgrounds in a short-baseline reactor neutrino experiment are determined in large part by the detector's surroundings, including the distance to the core, nearby spent fuel repositories, neutron backgrounds from the reactor and nearby experiments,  and cosmic ray induced backgrounds.  Also important are the cleanliness of the detector components, as well as the use of any background-reducing detection detection techniques.  The exact shape and magnitude of the backgrounds may vary widely and will only be known by conducting site-specific background surveys, detector material surveys, and detector simulations.

\begin{figure}[htbp]
   \centering
  \includegraphics[width=0.46\textwidth]{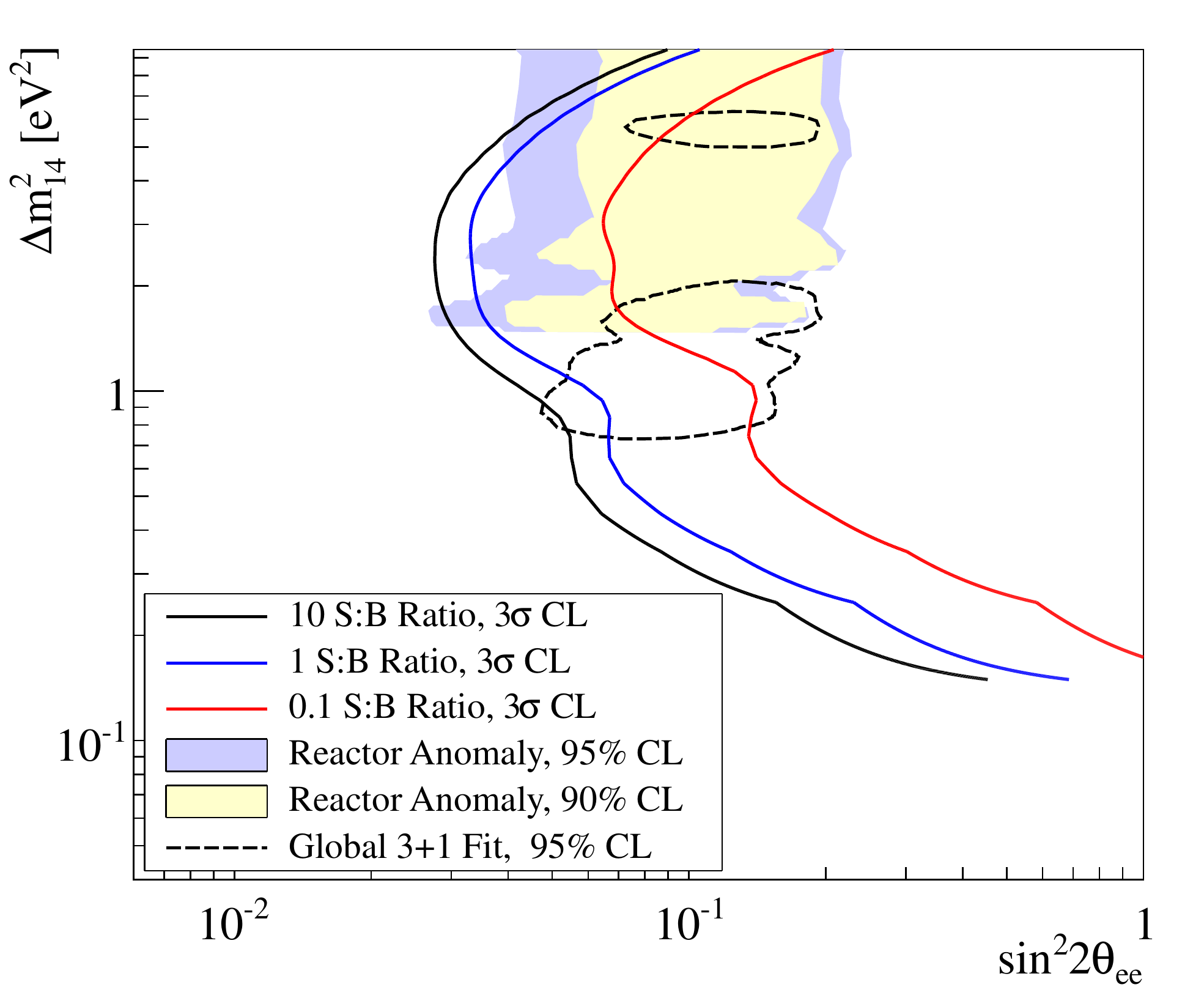}
\hspace{0.1cm}
  \includegraphics[width=0.46\textwidth]{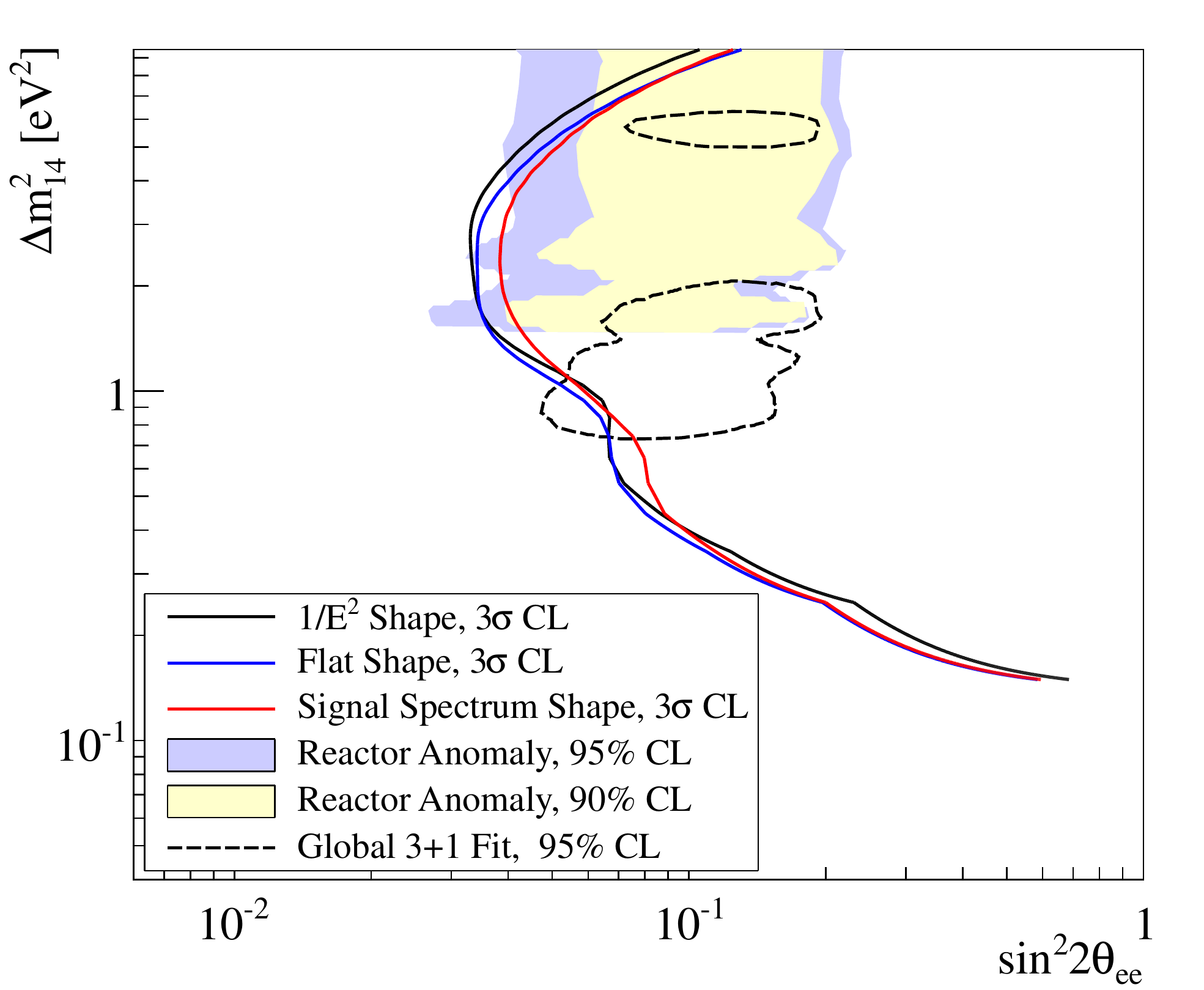}
   \caption{Variation in sensitivity of the default experiment with overall signal-to-background (left) and shape (right).  The default parameters used throughout the paper are a S:B ratio of 1:1 and a 1/E$^2$ spectral shape.  Backgrounds will play a critical role in the experimental sensitivity at a broad range of $\Delta m^2$ values, with the exact effect depending on the precise spectral shape as a function of energy and position.}
   \label{fig:Backgrounds}
\end{figure}

In place of detailed background surveys, the sensitivity of the default experiment to varying background conditions was investigated by varying the magnitude and shape of the input background spectrum in the $\chi^2$ calculation.  Signal-to-background ratios of 0.1, 1.0, and 10 were considered.  Changes in sensitivity resulting from variation of S:B ratio of the default experiment can be seen in Figure~\ref{fig:Backgrounds}.  The overall background rate clearly has a significant impact on experimental sensitivity at all values of $\Delta m^2$.  

As defined in the $\chi^2$ given in Equation~\ref{eq:chi2}, backgrounds are assigned a nuisance parameter that accounts for the uncertainty in the overall background normalization.  Wide variation of this parameter's associated systematic uncertainty has negligible effect on the sensitivity of the default experiment.

\begin{figure}[htbp]
   \centering
   \includegraphics[width=0.55\textwidth]{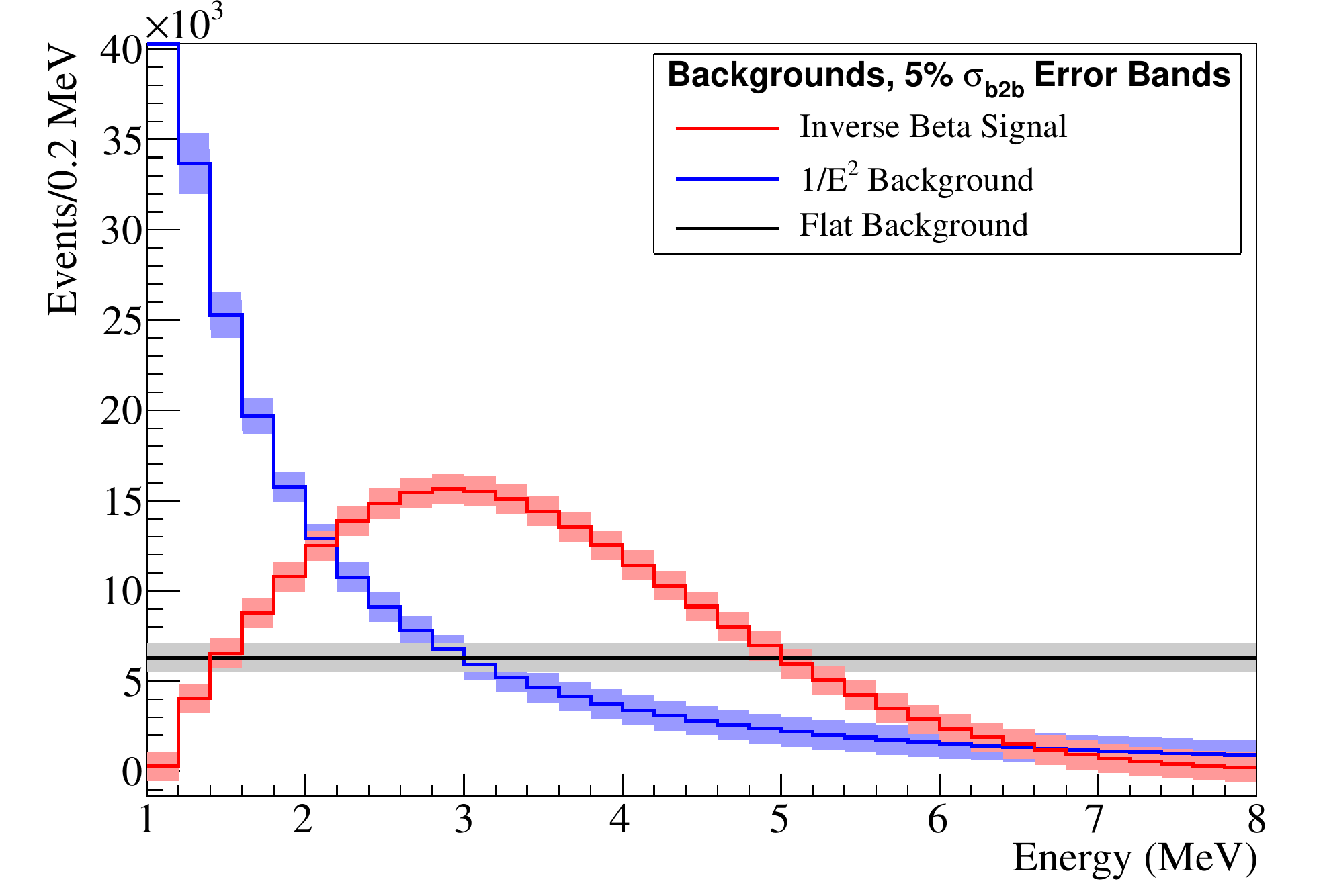} 
   \includegraphics[width=0.39\textwidth]{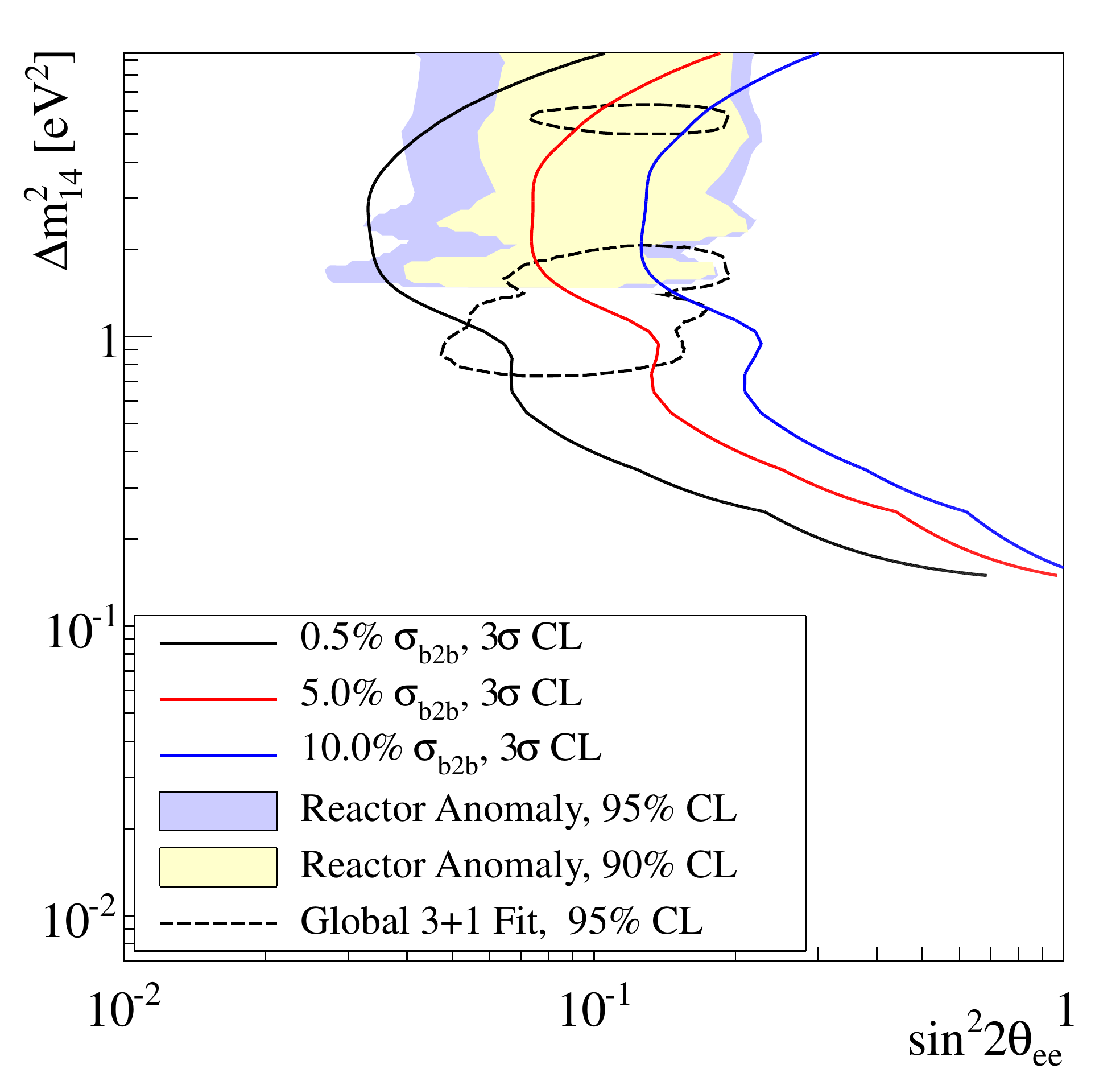} 
   \caption{Left: Example signal and background energy spectra.  A Signal:Background ratio of 1:1 as used in the default experiment is shown.  With a 1:1 ratio, the signal-shaped background spectrum is identical to the signal itself.  A 5\% uncertainty band indicates the effective range of uncorrelated variation allowed in the background spectrum by $\sigma_{b2b}$.  Right: Sensitivity of the default experiment to sterile neutrino oscillaions for various values of $\sigma_{b2b}$.  The significant variation in sensiivity with $\sigma_{b2b}$ underscores the importance of a precise knowledge of the background energy and position spectrum.}
   \label{fig:BkgShapes}
\end{figure}

Background spectral shapes were also varied in addition to overall background normalization. Three different background spectral shapes, pictured in Figure~\ref{fig:BkgShapes}, were used: the default 1/$E^2$ shape seen in previous short-baseline experiments~\cite{Bowden, ILL, Bryce}, a flat distribution commonly associated in deep-underground reactor $\nuebar$ experiments with fast neutron backgrounds, and a distribution identical to the signal distribution.  The results of these variations are also shown in Figure~\ref{fig:Backgrounds}.  The overall background normalization appears to have a much larger effect on sensitivity than the exact spectral shape, within the assumptions of this study. The same is also found to be true for the shape of the background position distribution.  

As discussed in Section~\ref{subsec:chisquare}, uncertainties in the background energy and position spectrum shapes, without site-specific background surveys,  conservatively accounted for with the addition of $\sigma_{b2b}$ to the denominator of the $\chi^2$.  To determine the effect of this uncorrelated spectral uncertainty on overall sensitivity, $\sigma_{b2b}$ was varied for the default experiment from 0.5\% to 10.0\%.  The effect of this parameter is illustrated in Figure~\ref{fig:BkgShapes}: each bin in energy and position is effectively allowed a free, uncorrelated fluctuation within a band determined by $\sigma_{b2b}$.  For a real experiment, the magnitude of $\sigma_{b2b}$ will be determined by measurements of background spectra during reactor-off periods, as well as measurements of low- and high-energy singles during reactor-on periods.  Shifting the detector radially may also help map out and constrain the spatial variation of the backgrounds.  The resultant change in sensitivity of the default experiment with variations in $\sigma_{b2b}$ are also shown in Figure~\ref{fig:BkgShapes}.  The experiment's sensitivity is highly dependent on $\sigma_{b2b}$ for all values of $\Delta m^2$.  Precise knowledge of the  background spectral shapes is clearly necessary for a definitive short-baseline oscillation experiment. 


\section{Detector Parameters}
\label{sec:detect}

\subsection{Target Mass and Efficiency}

The total neutrino interaction rate scales proportionally with target mass and detection efficiency.   Variation of target mass and efficiency have an identical effect on the sensitivity as variations in reactor power.  The target mass can be increased by increasing the cross-section of the detector, which is largely constrained by the available facility space, or by increasing the proton density of the target.  Efficiency is increased by increasing the fraction of prompt and delayed signal energy deposited in the active scintillating region of the detector or by relaxing signal selection cuts. 

\subsection{Detector Resolution and Analysis Binning}

Independent of a specific reactor-detector orientation and facility space constraints, the detector parameters that define the sensitivity of an experiment are the detector's energy and position resolution.  Good energy resolution is needed to both resolve the spectral distortions at a particular baseline and avoid washing out the oscillation at longer baselines, while position resolution allows for the observation of an oscillation as a function of distance. Previous short-baseline experiments have demonstrated the ability to attain sub-10\% energy resolution~\cite{Bugey, Rovno2}, while larger $\theta_{13}$ LS detectors have achieved resolutions of around 7-8\%~\cite{DBNIM}.  Acceptable position resolution can be obtained either through optical segmentation of the detector volume, or by using PMT charge topology information to reconstruct event positions in a larger one-zone detector.  The former technique has been demonstrated in a number of previous short-baseline experiments, such as Palo Verde and Bugey~\cite{Bugey, PaloVerde}, while the larger Gadolinium-loaded liquid scintillation detectors at Daya Bay, RENO, and Double Chooz have demonstrated successful position reconstruction with a resolution of 30~cm or better~\cite{DC, Bryce}.

\begin{figure}[htbp]
   \centering
  \includegraphics[trim=0.1cm 0.1cm 0.9cm 0.3cm, clip=true, width=0.46\textwidth]{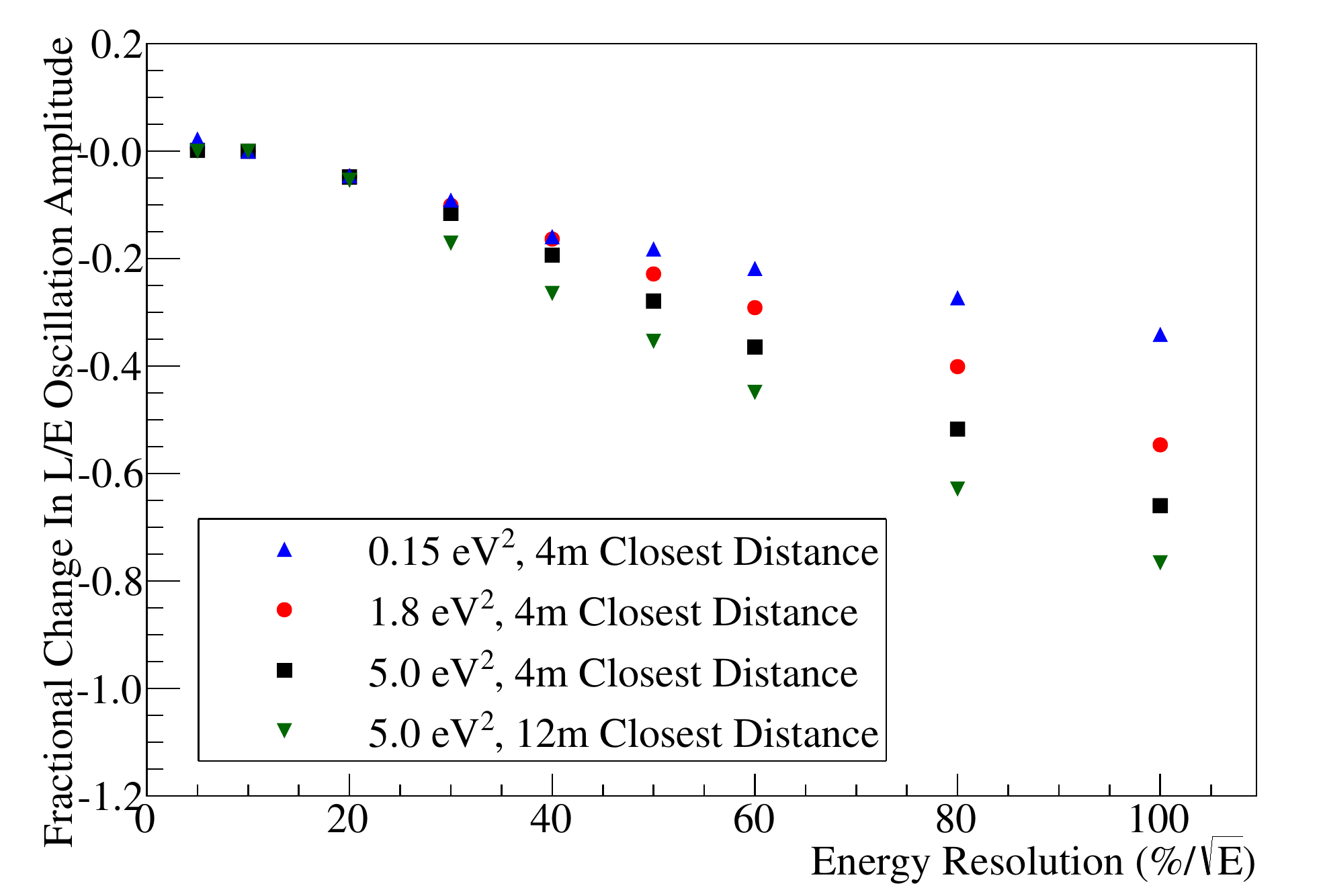}
\hspace{0.1cm}
  \includegraphics[trim=0.1cm 0.1cm 0.9cm 0.3cm, clip=true, width=0.46\textwidth]{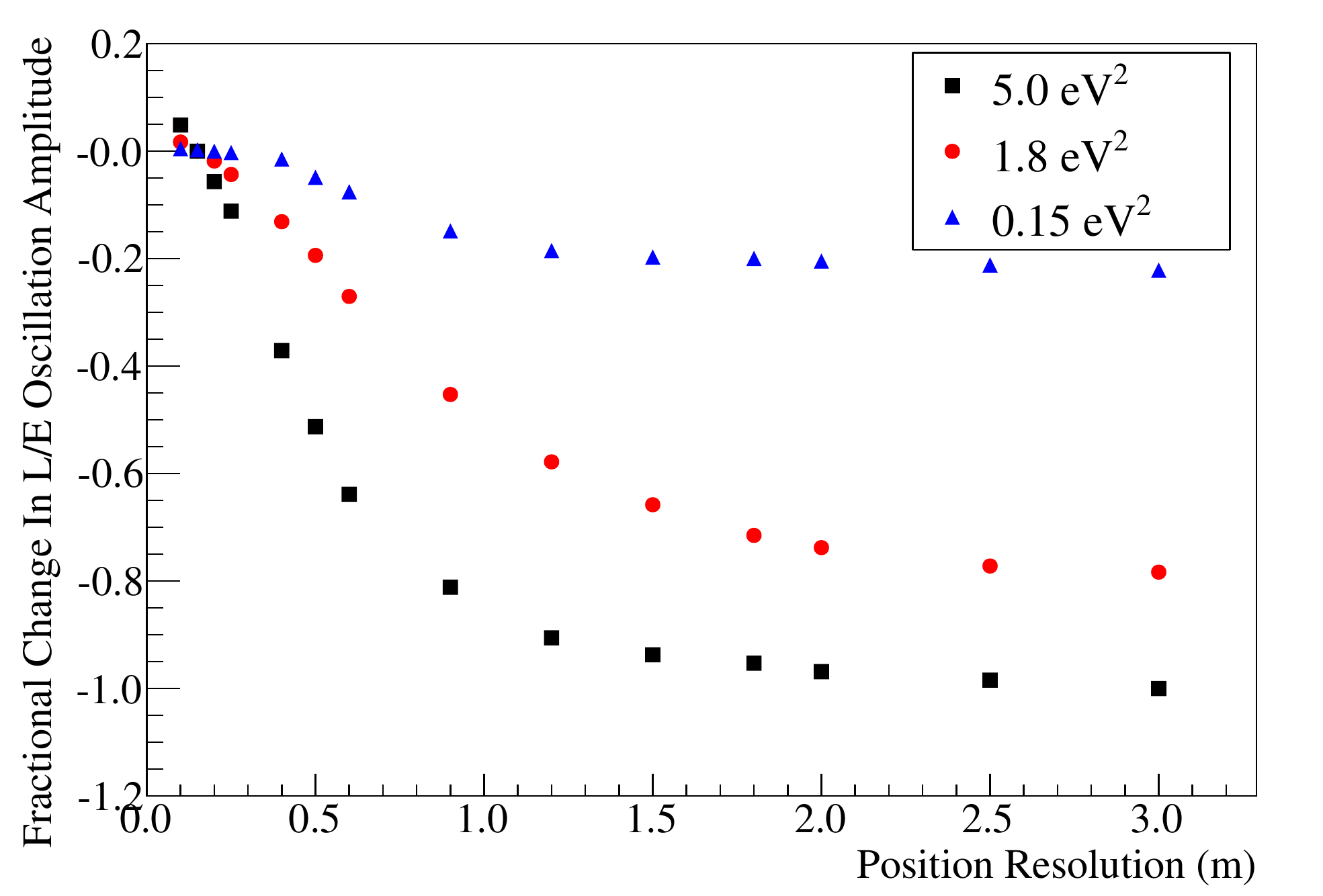}
   \caption{Fractional change in maximum oscillation sensitivity with changes in energy (left) and position (right) resolution for the default experiment at either 4~m or 12~m closest distance with for $\Delta m^2$ values of 0.15, 1.8, or 5.0 eV$^2$.  Fixed values of 10\% for energy and 15~cm for position are used when varying position resolution and energy resolution, respectively.  
}
   \label{fig:Res}
\end{figure}

Figure~\ref{fig:Res} demonstrates the effect of variations in detector resolution for the default experiment. The fractional change in amplitude of the measured L/E oscillation pictured in Figure~\ref{fig:LEOsc} is taken as the figure of merit.  The oscillation signature is increasingly washed out as the position and energy resolution decrease.  Attaining an energy resolution of better than 10\% will not provide significant gains in sensitivity for this range of $\Delta m^2$.   Position resolution approaching roughly 1/4 of the characteristic oscillation length at a particular $\Delta m^2$ is required to maximize sensitivity.

\begin{figure}[htbp]
   \centering
  \includegraphics[width=0.46\textwidth]{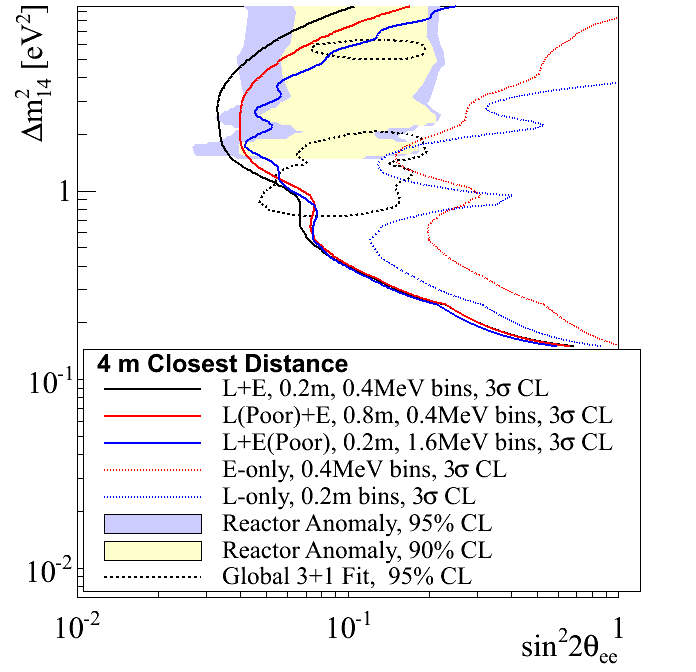}
\hspace{0.1cm}
  \includegraphics[width=0.46\textwidth]{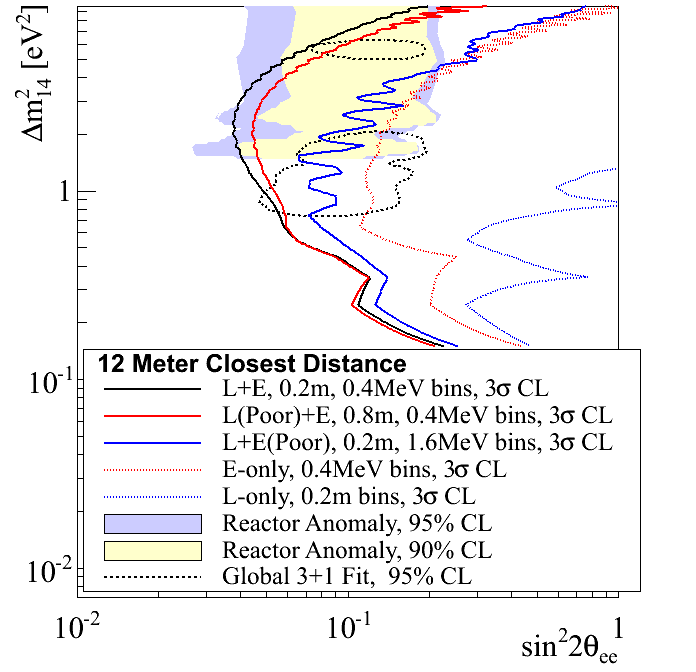}
   \caption{Variation in sensitivity to short-baseline oscillations with energy and position bin size for the default experiment at closest distances of 4~m (left) and 12~m (right).  Statistics are normalized to the level of the default experiment, 250,000 $\nuebar$/year.  Sensitivity is degraded, particularly at high $\Delta m^2$, as bin width is increased.  One can also see the differing contributions to the total sensitivity from position and energy information, and how this contribution changes with differing detector closest distance.}
   \label{fig:Ana}
\end{figure}

The effect of finite position and energy resolution can be approximated by varying the bin width in the $\chi^2$ analysis: the bin width provides an effective hard limit on the knowledge of exact event baselines and energies. 
The results of this approach are pictured in Figure~\ref{fig:Ana}.  The average $\nuebar$ detections in each bin are given in Table~\ref{tab:BinStats} to give a sense of the statistics provided by each binning.  It is clear that sensitivity to high $\Delta m^2$ values is degraded as bin width is increased.

\begin{table}[!htb]
  \noindent\makebox[\textwidth]{%
    \begin{tabular}{c|c|c|c} \hline
      \multicolumn{2}{c|}{Number of bins} & Events & \multirow{2}{*}{Comment} \\ \cline{1-2}
      Energy & Position & per bin &\\ \hline
      17 & 21 & 700 & Energy+Position Analysis, (0.4 MeV, 0.2 m) binning \\
      17 & 6 & 2450 & Energy+Position(Poor) Analysis, (0.4 MeV, 0.8 m) binning \\
      5 & 21 & 2380 & Energy(Poor)+Position Analysis, (1.6 MeV, 0.2 m) binning \\
      1 & 21 & 11900 & Position-only Analysis, 0.2 m binning \\
      17 & 1 & 14700 & Energy-only Analysis, 0.4 MeV binning \\ 
    \end{tabular}}
  \caption{Number of events per bin versus analysis binning.  250,000 total $\nuebar$ detections are expected with the default experiment.}
  \label{tab:BinStats}
\end{table}

Figure~\ref{fig:Ana} also illustrates qualitatively the individual contributions of energy and position information to the total sensitivity.  As the binning in energy or position, and thus the information available from energy or position, is reduced, sensitivity is reduced evenly, indicating that the two variables contribute roughly equally to the overall sensitivity of the default experiment.  This is a feature of our choice of a default detector and the resolutions considered. Figure~\ref{fig:Ana} also presents sensitivity for various binning choices of the same arrangement at 12~m closest distance.  For this arrangement, sensitivity is reduced much more quickly when energy binning is reduced, indicating that energy information provides most of the experimental sensitivity, and that energy resolution is of paramount importance for longer-baseline detectors.  

\section{Summary and Discussion}
\label{sec:summary}

\subsection{Key Experimental Parameters for an Experiment at Very Short Baselines}

In this paper we have studied an optimized very-short-baseline oscillation experiment utilizing an HEU core with the highest possible power and smallest core size.  To probe eV-scale mass-squared splittings, we consider a detector of at least 3~m at the shortest possible core distance with the highest possible efficiency, energy resolution of better than 10\%, and position resolution of better than 0.2~m.  Furthermore, a S/B ratio of at least 1:1 is desirable. The experiment may benefit from a second detector at longer baselines of 10-20\,m to improve sensitivity to the regions of small $\Delta m^2$.

\begin{figure}[htbp]
   \centering
   \includegraphics[trim=0.1cm 3cm 1cm 5.5cm, clip=true, width=0.95\textwidth]{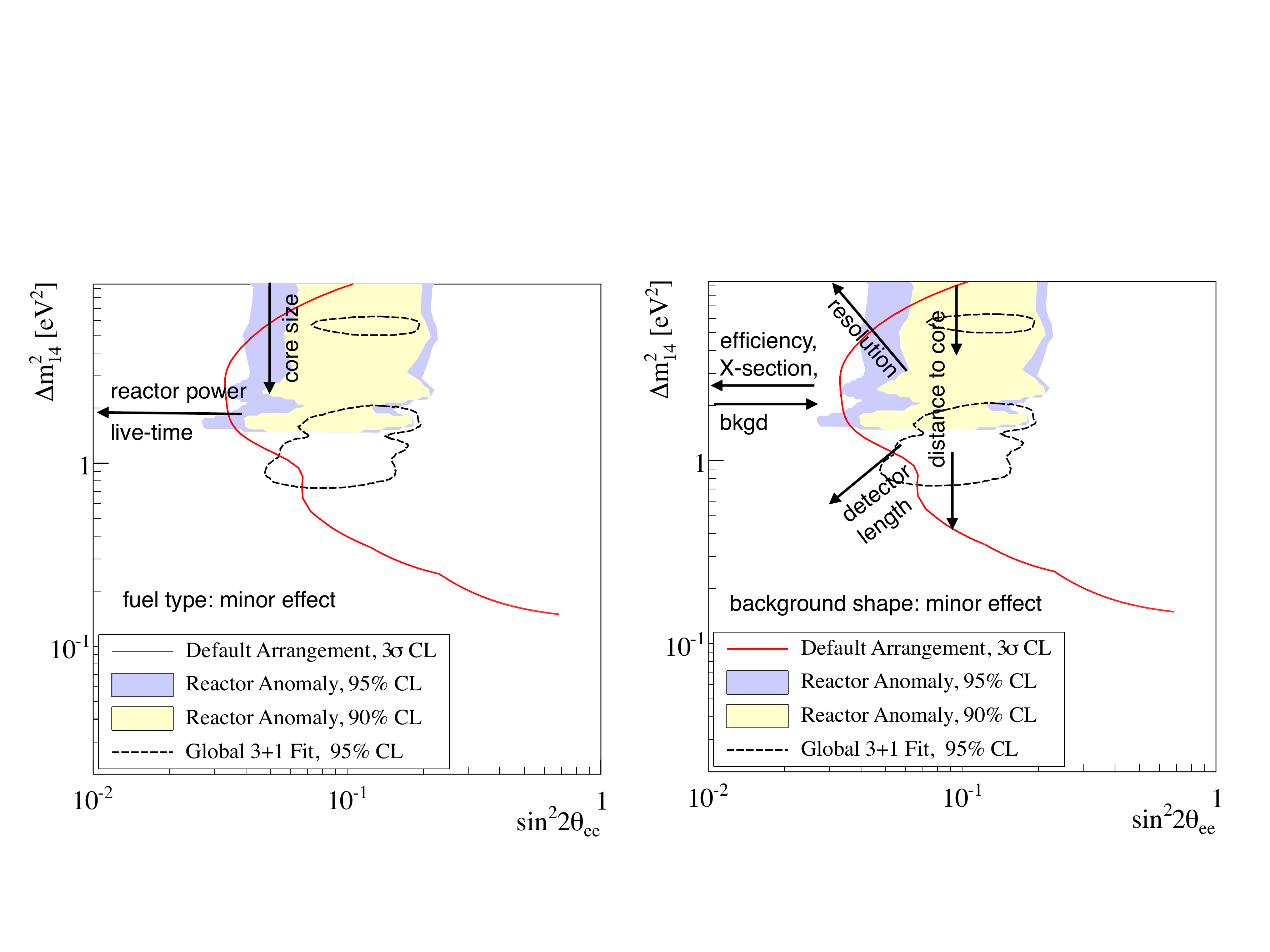} 
   \caption{Reactor and detector parameters relevant for covering the suggested parameter space.  These graphs indicate the direction in which the sensitivity curve moved when reactor (left) and detector (right) parameters are improved or adjusted.}
   \label{fig:summary}
\end{figure}

Figure~\ref{fig:summary} qualitatively summarizes the  impact of various reactor and detector parameters on the parameter space covered by very short baseline reactor $\nuebar$ experiments.  The relative important of these parameters is not weighted in this figure, and some experimental variables such as the precise core shape have only secondary effects on the experiment.


\begin{itemize}
  \item{\textbf{Total Statistics}: The per-bin statistical uncertainty can be a limiting quantity when searching for oscillations over numerous energy and position bins.  By utilizing a core with the highest possible power and a detector with the highest possible efficiency, cross-section, or proton density, one can maximize this quantity and considerably improve an experiment's sensitivity to short-baseline oscillations at all $\Delta m^2$ values.}
  \item{\textbf{Detector length}: A large detector length increases an experiment's ability to resolve oscillations with position in addition to spectral distortions in energy.  This effect can increase overall sensitivity at most $\Delta m^2$ values while extending the range of high sensitivity to lower values of $\Delta m^2$.}
  \item{\textbf{Detector-reactor distance}: The closest reactor-detector distance $r_{min}$ determines the $\Delta m^2$ range of highest sensitivity.  To achieve optimized sensitivity $r_{min}$  should be paired with a detector length $d$ of equal or larger magnitude to allow sampling of a broad range of the oscillation period.  In addition, statistics naturally increase as $r_{min}$ is decreased.}
  \item{\textbf{Detector resolution}: Oscillations at higher $\Delta m^2$ are only visible if resolutions and bin sizing are smaller than the oscillation itself.  Sensitivity to values of $\Delta m^2$ approaching this threshold are significantly reduced.}
  \item{\textbf{Background}: The S:B ratio is crucial for the success of the experiment. Small S:B ratios  make it  difficult to resolve oscillation effects above statistical background fluctuations and uncorrelated background uncertainties.  For a given S:B, various background spectral shapes have similar impact on the experiment. However, precise knowledge of the backgrounds and their distribution in energy and position are critical for the experiment's sensitivity and for demonstrating the observation of neutrino oscillation.}
\end{itemize}

Other variables, such as core size, shape, and fuel type will not likely drive experimental design at most mass splittings.  


\subsection{Energy Versus Energy+Position Measurements}
\label{subsec:Versus}

\begin{figure}[htbp]
   \centering
  \includegraphics[width=0.6\textwidth]{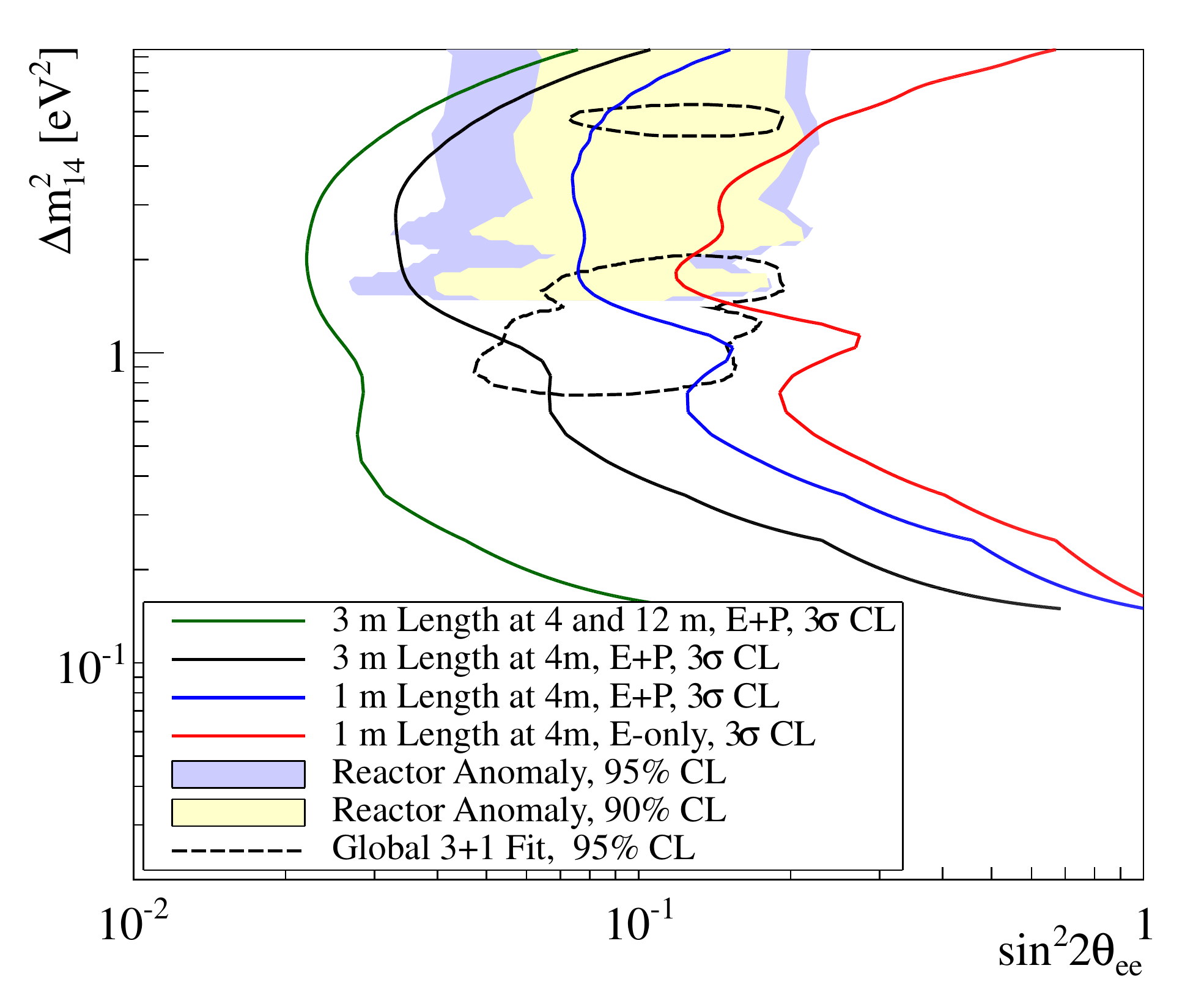}
  \caption{Comparison of sensitivities for oscillation experiments with energy-only or energy+position information.  All experimental parameters, including closest reactor-detector distance, are identical, except the total detector length.  
  Also pictured is an experiment utilizing an additional similar detector at a longer baseline of 12~m.  This extra detector greatly increases the available range of baselines, significantly increasing the experimental sensitivity at lower $\Delta m^2$ values.}
  \label{fig:Versus}
\end{figure}

Two classes of detectors have been proposed for very short baseline measurements at reactors.  The first consists of m$^3$-sized experiments 
that will look for oscillation-related distortions in the reactor $\nuebar$ energy spectrum.  The second class is larger multi-m$^3$ experiments with good energy and position resolution that are sensitive to both oscillations in position and energy.  The sensitivity of these two types of experiments can be compared using the default detector-reactor arrangement for each case while varying the total detector length: the energy-only experiment is defined to have a 1~m detector length, while the energy+position experiment is defined to have a 3~m detector length.  
A comparison of the sensitivities of three configurations is given in Figure~\ref{fig:Versus}: a 3~m detector with good position and energy resolution, a 1~m detector with good position and energy resolution, and a 1~m detector with no position reconstruction.  

Experiments with energy+position information provide significantly better sensitivity to oscillations than energy-only experiments.  Observing distortions of the energy spectrum at multiple baselines will lessen the effect of uncertainties in the energy spectrum shape.  If viewed at a single baseline, an energy spectrum distortion can be more easily described without oscillations by nuisance parameters for individual energy bins.  This problem is amplified if oscillations occur at high energies, where the reactor spectral uncertainties are larger.  In contrast, distortions from oscillation occur at different energy values for different baselines, an effect not easily neutralized with the comparatively rigid energy spectrum uncertainty nuisance parameters.  
Large differences also exist between experiments utilizing energy+position information in 1-m and 3-m long detectors. 
In addition to the difference in statistics,  the 1~m detector samples a shorter portion of the oscillation period as discussed in Section~\ref{subsec:Distance}.  
Even better sensitivity is achieved  by an experiment with two default detectors at distances of 4~m and 12~m as shown Figure~\ref{fig:Versus}.

\section{Conclusion}

New experiments at very short distances from reactors have the potential to make a precision measurement of the reactor antineutrino spectrum, resolve the ``reactor anomaly'',  and probe a large fraction of the sterile neutrino oscillation parameter space.  This paper explores the experimental variables of a short-baseline reactor experiment and resulting experimental sensitivity.  The reactor power, detector length, reactor-detector distance, energy and position resolutions of the detectors, and background to signal ratio are amongst the key parameters for such an experiment.  

Radially-extended detectors with good position and energy resolution provide a systematically robust approach to the search for neutrino oscillations at very short baselines. A highly-extended ($>$5m) detector or multiple detectors at different baselines may expand the sensitivity of the experiment to all relevant regions of $\Delta m^2$.  Furthermore, the use of moveable detectors would allow deployment at multiple radial locations, which may mitigate the possible effects of spatially varying backgrounds and will clearly demonstrate the effect of neutrino oscillations.

\acknowledgements
We thank Henry Band,  Jeff Cherwinka, and Randy Johnson for useful discussions and comments. This work was prepared with support from the Department of Energy, Office of High Energy Physics, under grants DE-FG02-95ER40896 and DE-FG02-84ER-40153, the University of Wisconsin, and the Alfred~P.~Sloan Foundation.

\bibliographystyle{revtex}
\bibliography{Reactorstudy}
\end{document}